\documentstyle[prd,aps,epsfig,tighten]{revtex}
\begin{document}
\preprint{                                                BARI-TH/309-98}
\draft
\title{			\hfill {\small\textsf{BARI-TH/309-98}}\\[2mm]
		Super-Kamiokande atmospheric neutrino data,\\
	   zenith distributions, and three-flavor oscillations	}
\author{        G.~L.~Fogli, E.~Lisi, A. Marrone, and G. Scioscia      }
\address{     Dipartimento di Fisica and Sezione INFN di Bari, 		\\
                  Via Amendola 173, I-70126 Bari, Italy			}
\maketitle
\begin{abstract}
We present a detailed analysis of the zenith angle distributions of 
atmospheric neutrino events observed in the Super-Kamiokande (SK) underground
experiment, assuming two-flavor and three-flavor oscillations (with one 
dominant mass scale) among active neutrinos. In particular, we calculate the 
five angular distributions associated to sub-GeV and multi-GeV $\mu$-like 
and $e$-like events and to upward through-going muons, for a total of 30 
accurately computed observables (zenith bins). First we study how such 
observables vary with the oscillation parameters, and then we perform a fit 
to the experimental data as measured in SK for an exposure of 33 kTy  (535
days). In the two-flavor mixing case, we confirm the results of the SK 
Collaboration analysis, namely, that $\nu_\mu\leftrightarrow\nu_\tau$ 
oscillations are preferred over $\nu_\mu\leftrightarrow\nu_e$, and that the 
no oscillation case is excluded with high confidence. In the three-flavor 
mixing case, we perform our analysis with and without the additional 
constraints imposed by the CHOOZ reactor experiment. In both cases, the 
analysis favors a  dominance of the $\nu_\mu\leftrightarrow\nu_\tau$ channel.
Without the CHOOZ constraints, the amplitudes of the subdominant 
$\nu_\mu\leftrightarrow\nu_e$ and $\nu_e\leftrightarrow\nu_\tau$ transitions 
can also be relatively large, indicating that, at present, current SK data 
do not exclude sizable $\nu_e$ mixing by themselves. After combining the 
CHOOZ and SK data, the amplitudes of the subdominant transitions are
constrained to be smaller, but they can still play a nonnegligible role both 
in atmospheric and other neutrino oscillation searches. In particular, we 
find that the $\nu_e$ appearance probability expected in long baseline 
experiments can reach the testable level of $\sim 15\%$. We also discuss 
earth  matter effects,  theoretical uncertainties, and various aspects of 
the statistical analysis. 
\end{abstract}

\section{Introduction}

The Super-Kamiokande (SK) water-Cherenkov experiment \cite{Open} has recently
confirmed \cite{SKSG,SKMG,EVID}, with high statistical significance, the
anomalous flavor composition of the observed atmospheric neutrino flux, as
compared with  theoretical expectations \cite{Hond,Bart}. The flavor anomaly
had been previously found in Kamiokande \cite{Ykam,Ymul}  and IMB
\cite{Yimb}, and later in Soudan2 \cite{Ysou}, but not in the  low-statistics
experiments NUSEX \cite{Nnus} and Fr{\'e}jus \cite{Nfre}.

The recent SK data have also confirmed earlier Kamiokande indications
\cite{Ymul} for a dependence of the flavor anomaly on the lepton zenith angle
\cite{SKMG,EVID}, which is correlated with the neutrino pathlength. The
features of this dependence are consistent with the hypothesis of  neutrino
oscillations, which represents the most natural (and perhaps  exclusive)
explanation of the data \cite{EVID}. The oscillation hypothesis  is also
consistent with other recent atmospheric neutrino data, namely, the 
finalized sample of Kamiokande upward-going muons \cite{KAUP}, the latest
muon and electron data from Soudan2 \cite{Soud}, and the samples of stopping
and through-going muons in MACRO \cite{MACR}.

The SK atmospheric $\nu$ measurements, which are described in detail in
several papers \cite{SKPU,SKSG,SKMG,EVID}, conference proceedings
\cite{Ke97,Kaji,Post}, and theses \cite{HaTh,KaTh,FlTh}, demand the greatest 
attention, not only for their intrinsic importance, but also for their 
interplay with other oscillation searches, including solar $\nu$ experiments 
\cite{SKSO} and long baseline oscillation experiments at reactors 
\cite{Cho1,Cho2,Palo} and accelerators \cite{Piet,MINO,KtoK,Beam,NUOI}.  In
this work, we contribute to these topics by performing a comprehensive, 
quantitative, and accurate study of the SK atmospheric $\nu$ data in the 
hypothesis of three-flavor mixing among active neutrinos. We also include, 
within the same framework, the recent data from the CHOOZ reactor experiment
\cite{Cho1,Cho2}.

More precisely, we consider the five SK angular distributions  associated to
sub-GeV and multi-GeV  $\mu$-like and $e$-like events  \cite{SKSG,SKMG} and
to upward through-going muons \cite{Kaji,Post}, for a  total of 30 accurately
computed observables (5+5+5+5+10 zenith bins).  First we study how such
observables vary with the oscillation parameters,  and then we fit them to
the experimental data as measured in SK for an  exposure of 33 kTy (535 days)
\cite{EVID,Kaji}.

In the two-flavor mixing case, we confirm the results of the SK 
Collaboration analysis, namely, that $\nu_\mu\leftrightarrow\nu_\tau$ 
oscillations are  preferred over $\nu_\mu\leftrightarrow\nu_e$, and that the 
no oscillation case is excluded with high confidence. In the three-flavor 
mixing case, we perform our analysis with and without the additional 
constraints imposed by the CHOOZ reactor experiment. In both cases, the 
analysis favors a  dominance of the $\nu_\mu\leftrightarrow\nu_\tau$ 
channel. Without the CHOOZ constraints, the amplitudes of the  subdominant
$\nu_\mu\leftrightarrow\nu_e$ and $\nu_e\leftrightarrow\nu_\tau$ transitions
can also be relatively large, indicating that, at present, current SK data do
not exclude sizable $\nu_e$ mixing by themselves. After combining  the CHOOZ
and SK data, the amplitudes of the subdominant  transitions  become smaller,
but we show that they can still play a  nonnegligible role both in
atmospheric, solar, and long baseline  laboratory experiments.

The plan of our paper is as follows. In Section~II we discuss  the 30 SK
observables used in the analysis, as well as the CHOOZ  measurement. In
Section~III we set the notation for our three-flavor  oscillation framework.
The two-flavor subcases are studied in Sec.~IV.  In Sec.~V we perform the
three-flavor analysis of SK data (with and without  the  CHOOZ constraints)
and discuss the results, especially those concerning $\nu_e$ mixing. In
Sec.~VI we study the implications of our analysis for the neutrino
oscillation phenomenology. We conclude our work in Sec.~VII, and devote
Appendixes~A and B to the discussion of  technical details related to our
calculations and to the statistical analysis.

Some of our previous results on two- and three-flavor oscillations  of solar 
\cite{3MSW,RVac,RMSW}, atmospheric \cite{Stat,3ATM,Marr,Zeni}, laboratory
\cite{AcRe,LBLE} neutrino experiments and their combinations 
\cite{Lisi,Sacr,Comp,Pert} will be often referred to in this work. In  
particular, the $3\nu$  analyses in \cite{3MSW,3ATM,Marr,Zeni,AcRe} 
summarize the pre-SK situation. However, we have tried to keep this paper as
self-contained as possible.

\section{Expectations and DATA}

In this Section we discuss expectations and data for five SK zenith
distributions: sub-GeV $e$-like and $\mu$-like,  multi-GeV $e$-like  and
$\mu$-like, and upward-going muons, with emphasis on some critical aspects of
both theory and data that are often neglected. Finally, we discuss the CHOOZ
reactor results.

\subsection{Zenith distributions of neutrinos and leptons}

A basic ingredient of any theoretical calculation or MonteCarlo simulation of
atmospheric $\nu$ event rates is the flux $\Phi$ of atmospheric neutrinos and
antineutrinos%
\footnote{We will often use the term ``neutrino'' loosely, to indicate 
	both $\nu$'s and $\overline\nu$'s. Of course, we properly
	distinguish $\nu$ from $\overline\nu$ in the input fluxes and
	in the calculations of cross sections and oscillation probabilities.}
as a function of the energy $E_\nu$ and of the zenith angle $\Theta$. The 
flux $\Phi(E_\nu,\Theta)$ is  unobservable in itself, and what is measured 
is the distribution of leptons  $\ell = \mu,e$ after $\nu$ interactions, as 
a function of the lepton energy $E_\ell$ $(<E_\nu)$ and lepton zenith angle 
$\theta$ $ (\neq \Theta)$.

In Fig.~1 we show the sum of theoretical $\nu$ and $\overline\nu$  fluxes, as
a function of the neutrino zenith angle $\Theta$, for selected values of the
energy $E_\nu$ (1, 10, and 100 GeV). The fluxes refer to the calculations of
\cite{Hond} (HKKM'95, solid lines) and \cite{Bart}  (AGLS'96, dots) without
geomagnetic corrections (so that the sign of  $\cos\Theta$ is irrelevant in
this figure). The upper and middle panels  refer to muon and electron
neutrinos, respectively, while the lower panel shows their ratio. Several
interesting things can be learned from this figure. For instance, the often
quoted value   $(\nu_\mu+\overline\nu_\mu)/(\nu_e+\overline\nu_e)\simeq 2$
for the muon-to-electron neutrino ratio clearly holds only for low-energy,
horizontal neutrinos. This ratio increases rapidly as the neutrino energy
increases and  as its direction approaches the vertical. In fact, both
$\nu_\mu$ and  $\nu_e$ fluxes decrease towards the vertical (see upper and
middle panels),  where the slanted depth in the atmosphere is reduced;
however,  $\nu_e$'s  are more effectively suppressed than $\nu_\mu$'s, due to
their different  parent decay chains. In addition, the greater the energy of
the parents,  the longer the decay lengths, the stronger the dependence of
the  $\nu_\mu/\nu_e$ flux ratio on the slanted depth and thus on the zenith 
angle $\Theta$. In other words, high-energy, vertical ``atmospheric $\nu$ 
beams'' are richer in $\nu_\mu$'s and, therefore, are best suited in 
searches for $\nu_\mu \to \nu_e$ oscillations, where initial $\nu_e$'s 
represent the ``background'' (see also \cite{Oyam}).  We anticipate that, in
fact, multi-GeV data are more effective than  lower-energy (sub-GeV) data in
placing bounds on  $\nu_\mu \to \nu_e$  transitions. Another consequence of
the non-flat $\nu_\mu/\nu_e$ ratio is  the appearance of distortions of the
zenith distributions that, although  related to {\em vacuum\/} neutrino
oscillation,  {\em do not\/} depend on  neutrino pathlength-to-energy ratio
$L/E_\nu$  \cite{Zeni}. Finally, notice  in Fig.~1 that the good agreement
at low energies between the two reported
calculations of  $\nu_\mu/\nu_e$  is somewhat
spoiled at high energies. This shows that the often-quoted uncertainty of 
$\pm 5\%$ for the $\nu_\mu/\nu_e$ ratio, which has been estimated in  detail
for {\em low energy, integrated fluxes\/} \cite{Rmue},  does not  necessarily
apply to high-energy or differential fluxes. Therefore, also  the $\mu/e$
lepton  event ratio  might suffer of uncertainties larger  than $\pm 5\%$ in
some energy-angle bins. Our empirical estimate of such  errors in the
statistical analysis is detailed in Appendix B. More  precise estimates of
the relative $\nu_\mu$ and $\nu_e$ flux uncertainties  are in progress
\cite{Gais,Giap}. A new, ab initio, fully three-dimensional  calculation of
the atmospheric neutrino flux \cite{Batt} is also expected  to shed light on
these issues.

\begin{figure}[t]
\begin{center}
\epsfig{bbllx=1truecm,bblly=8.8truecm,bburx=20truecm,bbury=25truecm,clip=,%
width=18truecm,file=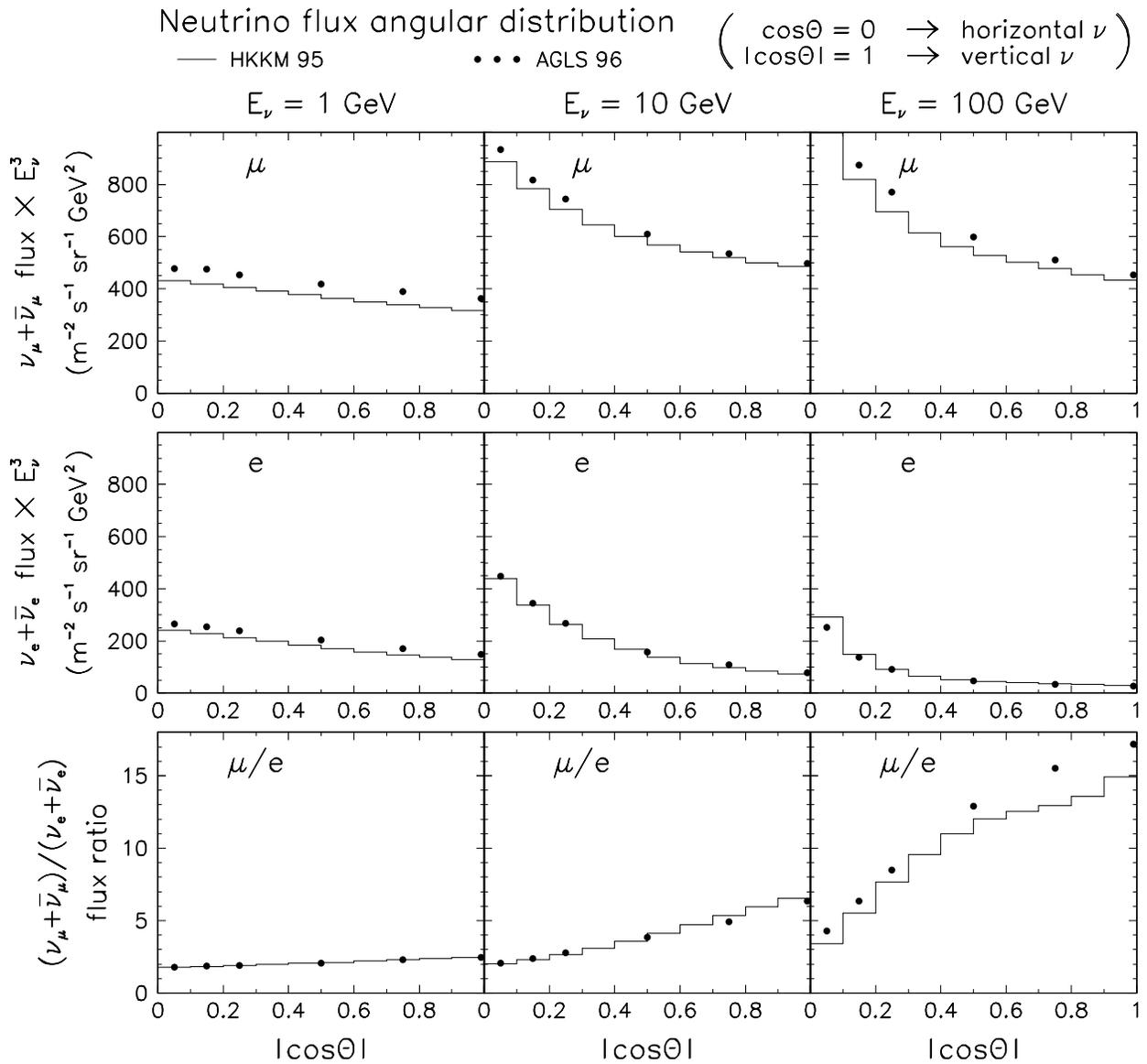}
\caption{Distributions of neutrino fluxes in terms of the neutrino
zenith angle $\Theta$, for three representative energies ($E_\nu=1$, 10, and
100 GeV). Upper panels: $\nu_\mu+\overline\nu_\mu$ flux. Middle panels:
$\nu_e+\overline\nu_e$ flux. Lower panels: 
$(\nu_\mu+\overline\nu_\mu)/(\nu_e+\overline\nu_e)$ flux ratio.
Solid lines and dots refer to the calculations in \protect\cite{Hond}
(HKKM'95) and \protect\cite{Bart} (AGLS'96), respectively. In this figure,
geomagnetic effects are not reported, and the fluxes are symmetric
under $\cos\Theta\to -\cos\Theta$. Notice the increase of the $\nu_\mu/\nu_e$
ratio with energy and for  $|\cos\Theta|\to 1$.}
\end{center}
\end{figure}

Concerning the overall uncertainty of the theoretical neutrino flux
normalization, it is usually estimated to be 20--30\%. Most  of the
uncertainty is associated to the primary flux of cosmic rays hitting the
atmosphere. It is important, however, to allow also for  energy-angle
variations of such normalization (e.g., the SK analysis  \cite{EVID} allows
the spectral index to vary within $\pm 0.05$). Our approach to this problem
is detailed in Appendix B. Here we want to  emphasize that valuable
information about the overall flux normalization  can be obtained
from more precise cosmic ray data from balloon experiments such as  BESS \cite{BESS}, CAPRICE
\cite{CAPR}, and MASS2 \cite{MASp,MASm}.  The BESS experiment has recently
reported a relatively low flux of cosmic  primaries \cite{BESS}, which, as we
will see, might represent a serious  problem for the oscillation
interpretation of the SK data  (see also  \cite{Gais,Giap}). On the other
hand,  the MASS2 experiment can also measure the flux of primary protons and
secondary muons {\em at the same time\/} \cite{Circ}, and might thus provide
soon an important calibration of the theoretical flux calculations
\cite{Circ,Cali}. Therefore, it is reasonable to expect, in a few years, a
reduction and a better understanding of the overall neutrino flux
uncertainty, with obvious benefits for the interpretation of the atmospheric
$\nu$ anomaly.

In order to obtain measurable quantities (e.g., the lepton zenith angle
distributions), one has to make a convolution of the neutrino  fluxes with
the differential cross sections and detection efficiencies  (see, e.g.,
\cite{3ATM,Comp,Pert}). We consider five zenith angle  distributions of
leptons: sub-GeV muons and electrons, multi-GeV muons and  electrons, and
upward through-going  muons. Concerning the calculation of  the first four
distributions, we have used the same technique used in  \cite{3ATM} for the
old Kamiokande multi-GeV distributions. This approach  makes use of the
energy distributions of parent {\em interacting} neutrinos \cite{Priv}.
Concerning upward through-going muons, we improve the
approach  used in \cite{Marr} by including the zenith dependence of the SK
muon  energy threshold as given in \cite{Post,HaTh}; we also use the same SK 
choice for the parton structure functions (GRV94 \cite{GRVD}, available  in
\cite{PDFL}) and muon energy losses in the rock \cite{Lohm,HaTh}. 
For all distributions (SG, MG, and UP$\mu$), we obtain a good
agreement with the corresponding distributions simulated by the SK
collaboration, as reported in Appendix~A (to which we refer the
reader for further details).

In short, we can compute five distributions of SK lepton  events as
a function of the zenith angle $\theta$, namely, sub-GeV  $\mu$-like and
$e$-like events (5+5 bins), multi-GeV $\mu$-like  and $e$-like events (5+5
bins), and upward through-going muons (10 bins),  for a total of 30
observables.%
\footnote{Our earlier calculations of event spectra for pre-SK experiments
	can be found in \protect\cite{3ATM,Marr,Pert,Comp}.}
Few other analyses report explicit calculations of sub-GeV and multi-GeV 
zenith distributions (see, e.g., \cite{Vall,Gonz}) or upward-going muon 
distributions (see, e.g., \cite{LiUP}) in agreement with the SK simulations.
Other authors perform detailed calculations but use  a reduced zenith
information, as that embedded, e.g., in the up-down lepton rate asymmetry
\cite{UPDO} (see, e.g., \cite{UDYa,Yasu}).

\subsection{SK Data: Total and differential rates}

The experimental data used in our analysis are reported in Table~I  and
Table~II, together with the corresponding expectations as taken from the SK
MonteCarlo simulations. The numerical values have been graphically reduced
from the plots in \cite{EVID,Kaji} and thus may be subject to slight
inaccuracies.

Table~I reports the zenith angle distributions of sub-GeV and  multi-GeV
events collected in the SK fiducial mass (22.5 kton) during 535  live days,
for a total exposure of 33 kTy. Fully and partially contained  multi-GeV
muons have been summed. Only single-ring events are considered. The
distributions are binned in five intervals  of equal width in  $\cos\theta$,
from $\cos\theta=-1$ (upward going leptons) to $\cos\theta=+1$  (downward
going leptons). The total number of events in the full solid angle  is also
given. The quoted uncertainties for the data points are statistical.  The 
statistical uncertainties associated to the SK MonteCarlo expectations
originate from the finite simulated exposure (10 years live time
\cite{Ke97}).%
\footnote{Systematic errors, not reported in Table~I,  are discussed in
	Appendix~B.}

Table~II reports the differential and total  flux of upward  through-going
muons as a function of the zenith angle. Data errors are  statistical only.
In this case, there is no statistical error for the SK  theoretical
estimates, which are derived from a direct calculation  \cite{Kaji,Post,HaTh}
and not from a MonteCarlo simulation.

It is useful to display the information in Tables~I and II in graphical form.
To this purpose, we take the central values of the   theoretical
expectations  in Tables~I and II as ``units of measure'' in  each bin. In
other words, all $\mu$ and $e$ event rates (either observed,  or calculated
in the  presence of oscillations) are normalized to their  standard (i.e.,
unoscillated) expectations $\mu_0$ and $e_0$.%
\footnote{This representation was introduced in \protect\cite{Stat} to 
	show those features of the atmospheric $\nu$ anomaly which are 
	hidden in the $\mu/e$ ratio and emerge only when $\mu$ and $
	e$ rates are separated.}
The following notations distinguish the various lepton samples:
$$
\begin{array}{ccl}
{\rm SG}e 	& = & \text{sub-GeV\ electrons}			\ ,\\
{\rm SG}\mu 	& = & \text{sub-GeV\ muons}			\ ,\\
{\rm MG}e 	& = & \text{multi-GeV\ electrons}		\ ,\\
{\rm MG}\mu 	& = & \text{multi-GeV\ muons}			\ ,\\
{\rm UP}\mu 	& = & \text{upward\ through-going\ muons}	\ ,\\
{\rm MC}    	& = & \text{theory\ (no\ oscillation)}		\ .
\end{array}
$$

\vspace*{1cm}

\begin{table}[t]
\caption{Super-Kamiokande  sub-GeV  and multi-GeV  $e$-like and $\mu$-like
atmospheric neutrino data,  compared with the corresponding MonteCarlo
simulations in different  zenith angle ($\theta$) bins \protect\cite{EVID}. 
The numerical values are graphically reduced from the plots in
\protect\cite{EVID,Kaji}. Units: Number of events. Experimental exposure:  33
kTy, corresponding to 22.5 kton fiducial mass  $\times$ 535 live days.
Simulated exposure: 10 yr  \protect\cite{Ke97}. Errors are statistical only.
The Super-Kamiokande MonteCarlo simulations refer to HKKM'95 neutrino fluxes 
\protect\cite{Hond}.}
\begin{tabular}{ccccccc}
 Event &Bin&$\cos\theta$&Observed  &MonteCarlo&Observed    &MonteCarlo  \\
 sample&No.&    range   &$e$-like events&$e$-like events
                        &$\mu$-like events&$\mu$-like events\\
\tableline
Sub-GeV  
    &1&$[-1.0,\,-0.6]$&$287\pm16.9$&$209\pm5.5$&$182\pm13.5$&$326\pm6.9$ \\
    &2&$[-0.6,\,-0.2]$&$231\pm15.2$&$206\pm5.5$&$225\pm15.0$&$316\pm6.8$ \\
    &3&$[-0.2,\,+0.2]$&$259\pm16.1$&$220\pm5.7$&$228\pm15.1$&$307\pm6.7$ \\
    &4&$[+0.2,\,+0.6]$&$227\pm15.1$&$216\pm5.6$&$264\pm16.2$&$308\pm6.7$ \\
    &5&$[+0.6,\,+1.0]$&$227\pm15.1$&$198\pm5.4$&$259\pm16.1$&$317\pm6.8$ \\
\cline{2-7}
&total&$[-1.0,\,+1.0]$&$1231\pm35.1$&$1049\pm12.4$&$1158\pm34.0$&$1574\pm15.2$\\
\tableline
Multi-GeV
    &1&$[-1.0,\,-0.6]$&$ 50\pm 7.1$&$ 37\pm2.3$&$ 64\pm 8.0$&$114\pm4.1$ \\
    &2&$[-0.6,\,-0.2]$&$ 56\pm 7.5$&$ 51\pm2.7$&$ 75\pm 8.7$&$132\pm4.4$ \\
    &3&$[-0.2,\,+0.2]$&$ 70\pm 8.4$&$ 62\pm3.0$&$136\pm11.7$&$173\pm5.0$ \\
    &4&$[+0.2,\,+0.6]$&$ 74\pm 8.6$&$ 52\pm2.8$&$142\pm11.9$&$139\pm4.5$ \\
    &5&$[+0.6,\,+1.0]$&$ 40\pm 6.3$&$ 34\pm2.2$&$114\pm10.7$&$111\pm4.0$ \\
\cline{2-7}
&total&$[-1.0,\,+1.0]$&$290\pm17.0$&$236\pm5.9$&$531\pm23.0$&$669\pm9.9$ 
\end{tabular}
\end{table}

\vspace*{3cm}


\begin{table}[b]
\caption{Super-Kamiokande 535 day data on upward through-going muon fluxes,
compared with the corresponding theoretical calculations in different zenith
angle ($\theta$) bins \protect\cite{Kaji}. Units:
$10^{-13}$~cm$^{-2}$~s$^{-1}$~sr$^{-1}$.  The calculated muon fluxes, taken
from \protect\cite{Kaji,Post},  refer to HKKM'95 neutrino fluxes
\protect\cite{Hond}, GRV'94 DIS structure functions  \protect\cite{GRVD}, and
Lohmann muon energy losses in the rock \protect\cite{Lohm}. Errors are
statistical only. The numerical values are graphically reduced from the plots
in \protect\cite{Kaji,Post}.}
\begin{tabular}{cccc}
Bin 	& $\cos\theta$ 		&     Observed 		& Theoretical 	\\
No. 	&   range        	&   $\mu$ flux  	& $\mu$ flux    \\
\tableline
  1	&  $[-1.0,\,-0.9]$	& $ 1.03 \pm 0.18 $	& 1.25		\\
  2	&  $[-0.9,\,-0.8]$	& $ 1.16 \pm 0.18 $	& 1.38		\\
  3	&  $[-0.8,\,-0.7]$	& $ 0.90 \pm 0.17 $	& 1.46		\\
  4	&  $[-0.7,\,-0.6]$	& $ 1.62 \pm 0.22 $	& 1.57		\\
  5	&  $[-0.6,\,-0.5]$	& $ 1.31 \pm 0.18 $	& 1.67		\\
  6	&  $[-0.5,\,-0.4]$	& $ 1.57 \pm 0.20 $	& 1.78		\\
  7	&  $[-0.4,\,-0.3]$	& $ 1.59 \pm 0.21 $	& 1.93		\\
  8	&  $[-0.3,\,-0.2]$	& $ 2.20 \pm 0.25 $	& 2.18		\\
  9	&  $[-0.2,\,-0.1]$	& $ 2.73 \pm 0.28 $	& 2.52		\\
 10	&  $[-0.1,\,-0.0]$	& $ 3.42 \pm 0.31 $	& 3.03		\\
\tableline
total	&  $[-1.0,\,-0.0]$	& $ 1.75 \pm 0.07 $	& 1.88		
\end{tabular}
\end{table}%

\begin{figure}[t]
\begin{center}
\epsfig{bbllx=2.0truecm,bblly=8.8truecm,bburx=20truecm,bbury=23truecm,clip=,%
width=18truecm,file=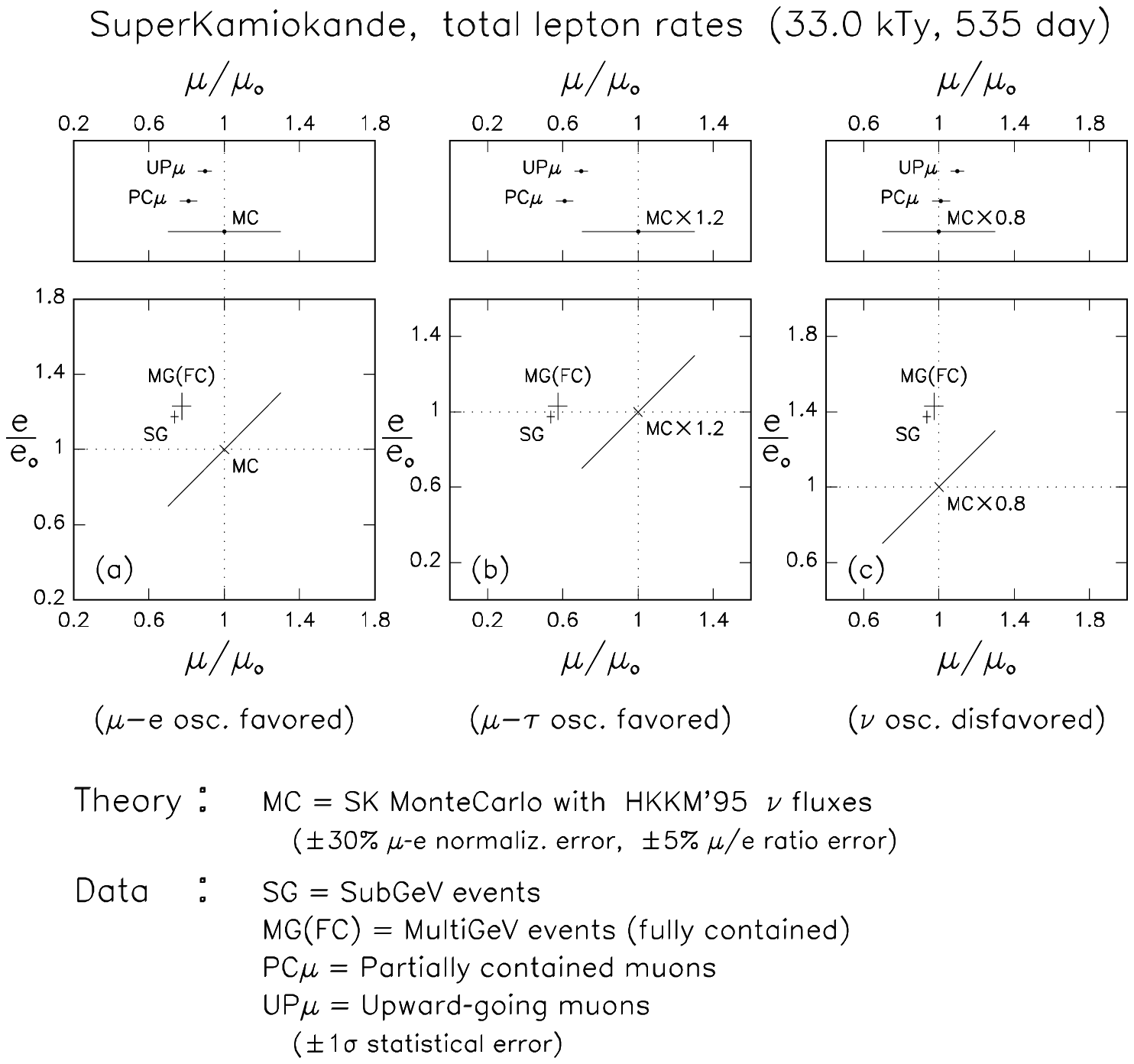}
\caption{Super-Kamiokande data on total lepton events,  compared with
their theoretical predictions.  The numbers of $e$-like and $\mu$-like events
are normalized to the central values of the corresponding MonteCarlo (MC)
simulations $e_0$ and $\mu_0$ \protect\cite{EVID}, obtained with HKKM'95
fluxes  \protect\cite{Hond} (as reported in Table~I). Error bars of
experimental data are statistical; slanted error bars of MC represent  $\pm
30\%$ systematics in the common $\mu_0,e_0$ normalization and $\pm5\%$
systematics in the $\mu_0/e_0$ ratio. Left panel: Default MC (favors
$\nu_\mu\to\nu_e$ oscillations). Middle panel: MC$\times 1.2$ (favors
$\nu_\mu\to\nu_\tau$ oscillations).  Right panel: MC$\times0.8$ (disfavors
$\nu$ oscillations).  See the text for details.}
\end{center}
\end{figure}

Figure~2 compares theory and data for the total lepton rates. The MG$\mu$
data sample is further divided into fully contained (FC) and  partially
contained (PC) events (notice that MG$e$ events are all FC). The  total
number of lepton events are displayed in the plane $(\mu/\mu_0,e/e_0)$, so
that the standard expectations (MC)  correspond to the point $(1,1)$ for 
each data sample.  The UP and PC muons have no electron counterpart and are 
shown in the single variable $\mu/\mu_0$ (upper strips). We attach a
$\pm30\%$  common uncertainty to the MC muon and electron rates  (large
slanted error bar), and allow for a $\pm 5\%$ uncertainty in the  $\mu/e$
ratio (small slanted error bar).  The three subfigures 2(a), 2(b), and 2(c)
correspond to three choices for the theoretical predictions (MC):  (a)
Standard MC expectations;  (b) MC rates multiplied by 1.2; and  (c) MC rates
multiplied by 0.8.

Figure~2(a) clearly shows that, with respect to the standard MC, SK  observes
a deficit of muons (stronger for low-energy SG$\mu$ data and weaker  for
high-energy UP$\mu$  data) and, at the same time, an excess of electrons 
(both in the SG and MG samples).  In the hypothesis of neutrino oscillations,
these indications would favor $\nu_\mu\to\nu_e$ transitions, with a mass 
square difference low enough to give some energy dependence to the muon 
deficit. As far as {\em total\/} SK rates are concerned, this is a perfectly 
viable scenario.

\begin{figure}[t]
\begin{center}
\epsfig{bbllx=2.0truecm,bblly=9.4truecm,bburx=20truecm,bbury=19.4truecm,clip=,%
width=18truecm,file=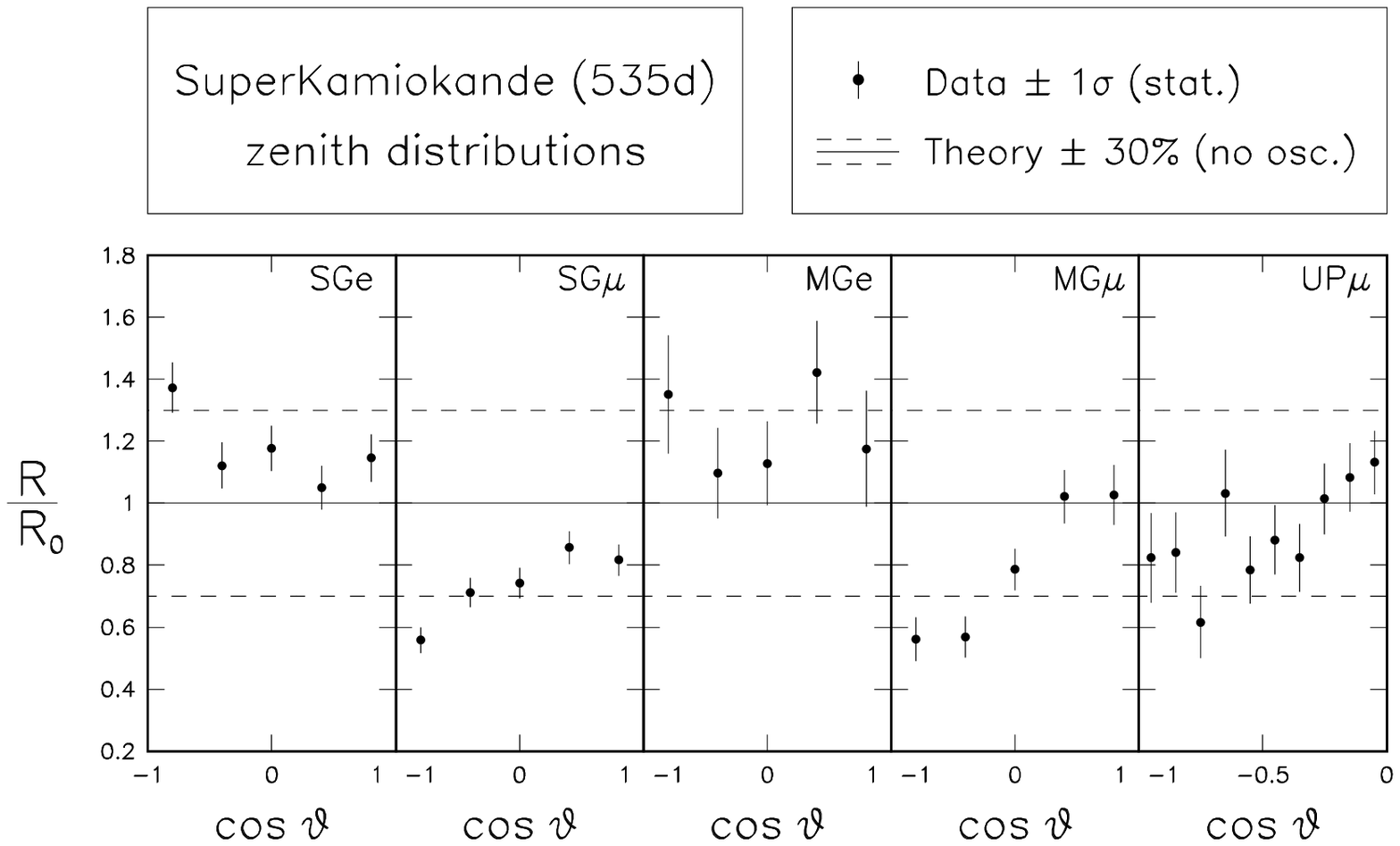}
\caption{Super-Kamiokande distributions of lepton events in terms
of the lepton zenith angle $\theta$ ($\cos\theta=-1,0,1$ correspond
to upgoing, horizontal, downgoing leptons).  From left to right:  sub-GeV
electrons (SG$e$) and muons (SG$\mu$), multi-GeV electrons (MG$e$) and muons
(MG$\mu$), and upward through-going muons UP$\mu$. In each bin, the observed
rate $R$ is divided by the expected rate $R_0$ in the absence of
oscillations,  as taken from Tables~I and II. 
Therefore, in this plot Theory=1 with
$\pm 30\%$ normalization error, and the deviations of Data (dots) from the
flat theoretical expectations show the zenith anomaly at glance.}
\end{center}
\end{figure}

Figure~2(b) shows how the previous picture changes when one allows  for an
overall increase of the MC expectations (say, $+20\%$). The relative excess
of electrons disappears, while the muon deficit is enhanced.  This situation
is consistent with $\nu_\mu\to\nu_\tau$ transitions, which leave the electron
rate unaltered. The current SK data link rather strongly $\nu_\mu\to\nu_\tau$
oscillations to an overall increase of the MC expectations: indeed, a good
fit to the SG and MG data samples requires a MC ``renormalization'' by a
factor $\sim 1.16$ \cite{EVID}.  Although this factor is acceptable at
present, it might not be so in the  future, should the MC predictions become
more constrained. In particular,  if the recent BESS indications \cite{BESS}
for a relatively low flux of cosmic primaries were confirmed, then one should
rather {\em decrease\/} the MC expectations \cite{Gais,Giap}.

Figure~2(c) shows the effect of a MC decrease by 20\%. In this case one would
observe no deficit of muons (and even an excess of UP$\mu$ events) and a
$\sim 40\%$ excess of electrons, which cannot be obtained in any known
oscillation scenario.

The discussion of Fig.~2 shows that: 
(i)   Theory (no oscillation) and data disagree,  even allowing for a MC 
      renormalization; 
(ii)  The oscillation interpretation depends sensitively on the size
      of the renormalization factor; 
(iii) If this factor turns out to be $<1$, the oscillation
      hypothesis is jeopardized; and
(iv)  It is thus of the utmost importance to calibrate and constrain the 
      theoretical neutrino flux calculations \cite{Hond,Bart} through cosmic 
      ray balloon experiments such as BESS \cite{BESS}, and especially 
      through simultaneous measurements of primary and secondary charged 
      particles as in the forthcoming CAPRICE and MASS2 analyses \cite{Circ}.
All this information would be lost if the popular ``$\mu/e$ double ratio'' 
were used. Another piece of information that would be hidden by the double 
ratio is the fact that, in Fig.~2, the SG and MG data points appear to be 
very close to each other in the $(\mu/\mu_0,e/e_0)$ plane, while it was not 
so in Kamiokande \cite{Stat}.  We are, however,
unable to trace the source of such a 
difference (which is  independent of the MC normalization) between SK and
Kamiokande.

Since the total lepton rate information is subject to the above ambiguities,
one hopes to learn more from {\em differential rates\/} and,  in particular,
from the zenith distributions of electrons and muons. These  distributions
are shown in Fig.~3 where, again, the rates $R$ have been  normalized in each
bin to the central values of their expectations $R_0$  (from Tables~I and
II). Therefore, the no oscillation case corresponds to ``theory~=~1'' in this
figure, with an overall normalization error that we set at $\pm 30\%$.
Deviations of the data samples (dots with error bars) from the standard
(flat) distribution are then immediately recognizable. The electron samples
(SG$e$ and MG$e$) do not show any significant deviation from a flat shape,
with the possible exception of a slight excess of upward-going
($\cos\theta\to -1$) SG electrons. On the other hand, all the muon samples
show a significant slope in the  zenith distributions, especially for
multi-GeV data.%
\footnote{It must be said that, in general, one cannot expect very strong 
	zenith deviations in the SG data distribution,  since the 
	neutrino-lepton scattering angles are typically large at low energies
	($60^\circ$, on average) and therefore the flux of leptons is more 
	diffuse in the solid angle.} 
Most of this work is devoted to understand how well two- and three-flavor 
oscillations of active neutrinos can explain these features of the SK angular
distributions.

We mention that additional SK measurements can potentially  corroborate the
neutrino oscillation hypothesis,  namely, the  stopping-to-passing  ratio of
upward-going muons \cite{Kaji,Post},  neutral-current enriched event samples
\cite{Kaji,UPNC,Pion,PiNC,Mrin},  and azimuth (east-west) distributions of
atmospheric  $\nu$ events \cite{Kaji,GEOM}. Such preliminary data will be
considered in a future work.

Finally,  the SK data themselves could be used for a  self-calibration of the
overall neutrino flux normalization. In particular,  one should isolate a
sample of high-energy, down-going leptons with  directions close to the
vertical, so that the corresponding parent neutrinos  would be characterized
by a pathlength, say, $L\lesssim 50$ km and by an  energy  $E_\nu\gtrsim
10$--20 GeV. Then, for a neutrino mass difference  $\Delta m^2$ smaller than
$10^{-2}$ eV$^2$ \cite{EVID}, the oscillating  phase $\propto \Delta m^2
L/E_\nu$ would also be small, and the selected  sample could be effectively
considered as {unoscillated\/}, thus providing  a model-independent
constraint on the absolute  lepton rate and on the neutrino flux. The present SK
statistics for strictly down-going, high-energy leptons, is not yet adequate
to such a calibration. We will come back to this issue in the following.

\subsection{CHOOZ results}

The CHOOZ experiment \cite{Cho1} searches for possible $\overline \nu_e$
disappearance by means of a detector placed at $L\simeq 1$ km from two
nuclear reactors with a total thermal power of 8.5 GW. With an average value
of $L/E_\nu\sim 300$ km/GeV, it is able to explore the 
$\overline\nu_e\to\overline\nu_e$ oscillation channel down to $\sim 10^{-3}$ 
eV$^2$ in the neutrino mass square difference, improving by about an order  of
magnitude previous reactor limits \cite{AcRe}. The sensitivity to  neutrino
mixing is at the level of a few percent, being mainly limited  by systematic
uncertainties in the absolute reactor neutrino flux. The  ratio of observed
to expected neutrino events is $0.98\pm0.04\pm0.04$,  thus placing strong
bounds on the electron flavor disappearance \cite{Cho1}.  The CHOOZ limits
have been recently retouched (weakened) \cite{Cho2} as  a result of the
unified approach to confidence level limits proposed in  \cite{Cous}.

The impact of CHOOZ  for atmospheric oscillation searches and for their
interplay with solar neutrino oscillations \cite{CHde} has  been widely
recognized (see, e.g.,  \cite{Lisi,Yasu,Barb,Barg,Mina,Bima}).  Earlier
studies of the interplay between reactor, atmospheric, and solar  neutrino
experiments can be found in \cite{3MSW,3ATM,AcRe,Comp,Pert}.

Given their importance, we have performed our own reanalysis of  the CHOOZ
data in order to make a proper SK+CHOOZ combination. We use the  $\overline
\nu_e+p\to e^++n$ cross section as in our previous works  \cite{AcRe}, and
convolute it with the reactor neutrino energy spectrum  \cite{Cho1} in order
to obtain the positron rate. The expected rate is  then compared with the
data \cite{Cho1,Cho2} through a $\chi^2$ analysis.  We have checked that, in
the case of two-family oscillations, we obtain  with good accuracy the
exclusion limits shown in \cite{Cho2}. Our  CHOOZ reanalysis will be
explicitly presented in Section~IV. The CHOOZ bound counts as one additional
constraint (the observed positron rate); therefore, the global SK+CHOOZ
analysis represents  a fit to 30+1 observables.

\section{Three-flavor framework and two-flavor subcases}

In this Section we set the convention and notation used in the  oscillation
analysis. We consider three-flavor mixing among active  neutrinos:%
\footnote{Oscillations into sterile neutrinos (see, e.g., 
	\protect\cite{Gonz,Smst,List,Yast} and references therein) 
	are not considered in this paper. }
\begin{equation}
\left(\begin{array}{c}
			\nu_e   \\ 
			\nu_\mu \\ 
			\nu_\tau
\end{array}\right)
			=
\left(\begin{array}{ccc}
		U_{e1} 		& U_{e2} 	& U_{e3} 	\\
		U_{\mu1} 	& U_{\mu2} 	& U_{\mu3} 	\\
		U_{\tau1} 	& U_{\tau2} 	& U_{\tau3} 
\end{array}\right)
\left(\begin{array}{c}
			\nu_1   \\ 
			\nu_2   \\ 
			\nu_3
\end{array}\right)
\ ,
\end{equation}
with
\begin{equation}
{\rm mass}(\nu_1,\nu_2,\nu_3)\equiv (m_1,m_2,m_3)\ .
\end{equation}

It is sometimes useful to parametrize the mixing matrix $U_{\alpha i}$
in terms of three mixing angles, $\omega$, $\phi$, and $\psi$:
\begin{equation}
	U_{\alpha i} =
\left(\begin{array}{ccc}
	c_\phi c_\omega & 
	c_\phi s_\omega &  
	s_\phi 
\\
	-s_\psi s_\phi c_\omega - c_\psi s_\omega & 
	-s_\psi s_\phi s_\omega + c_\psi c_\omega & 
	s_\psi c_\phi 
\\
	-c_\psi s_\phi c_\omega + s_\psi s_\omega & 
	-c_\psi s_\phi s_\omega - s_\psi c_\omega & 
	c_\psi c_\phi 
\end{array}\right)\ ,
\end{equation}
where $c=\cos$, $s=\sin$, and we have neglected a possible CP violating 
phase that, in any case, would be unobservable in our framework. The mixing
angles $(\omega,\phi,\psi)$ are also indicated as
$(\theta_{12},\theta_{13},\theta_{23})$ in the literature.

While three-flavor oscillation probabilities are trivial to be  computed in
vacuum (i.e., in the ``atmospheric part'' of the neutrino  trajectory), 
refined calculations are needed to account also for matter  effects in the
Earth. As in our previous works \cite{3ATM,Pert}, we solve  numerically the
neutrino evolution equations for any neutrino trajectory,  taking into
account the corresponding electron density profile in the Earth. Our computer
programs are designed to compute the neutrino and antineutrino oscillation
probabilities for {\em any possible} choice  of values for the neutrino
masses $(m_1,m_2,m_3)$ and mixing  angles $(\omega,\phi,\psi)$.

\subsection{Neutrino masses}

A complete exploration of the three-flavor neutrino parameter space  would be
exceedingly complicated. Therefore, data-driven approximations are  often
used to simplify the analysis \cite{Sacr}.  As in our previous works 
\cite{3MSW,3ATM,AcRe,Comp}, we use the following hypothesis about neutrino
square mass differences:
\begin{equation} |m^2_2-m^2_1| \equiv \delta m^2 \ll m^2 \equiv |m^2_3 -
m^3_2|\ , \label{appr} \end{equation}
i.e., we assume that one of the square mass differences  ($\delta m^2$) is
much smaller than the other $(m^2)$, which is the one probed by atmospheric
neutrino experiments (and accelerator or reactor experiments as well
\cite{AcRe}). The small square mass difference is then presumably associated
to solar neutrino oscillations \cite{3MSW}.

Notice that the above approximation involves squared mass differences and not
the absolute masses (which cannot be probed in oscillation searches). In
particular, Eq.~(\ref{appr}) simply states that there is a  ``lone neutrino''
$\nu_3$, and a ``neutrino doublet'' $(\nu_1,\nu_2)$, the  doublet mass
splitting being much smaller than the mass gap with the lone  neutrino.
However, Eq.~(\ref{appr}) can be fulfilled with either  $m_3>m_{1,2}$ or
$m_3<m_{1,2}$. These two cases are not entirely equivalent  when matter
effects are taken into account, as shown in \cite{3ATM}.  However, the
difference is hardly recognizable in the current atmospheric  $\nu$
phenomenology \cite{3ATM,Yasu}.  For simplicity, in this paper we refer only
to the case $m_3>m_{1,2}$, i.e., to  a ``lone''  neutrino $\nu_3$ being the
heaviest one.

As far as $m^2\gtrsim 10^{-4}$ eV$^2$ and $\delta m^2 \ll 10^{-4}$ eV$^2$,
atmospheric neutrino oscillations depend effectively only on $m^2$. However,
for larger values of $\delta m^2$ the approximation (\ref{appr}) begins to
fail, and subleading, $\delta m^2$-driven oscillations can affect the
atmospheric $\nu$ phenomenology. We will briefly comment on subleading
effects in Sec.~VI~C. Earlier discussions of such effects in solar
and atmospheric neutrinos can be found in \cite{Pert}.

\subsection{Neutrino mixing}

Under the approximation (\ref{appr}) one can show that  CP violating effects
are unobservable, and that the angle $\omega$ can be rotated away in the
analysis of atmospheric neutrinos (see \cite{3ATM} and references therein).
In other words, atmospheric $\nu$ experiments  do not probe the mixing
$\omega=\theta_{12}$ associated with the  quasi-degenerate doublet
$(\nu_1,\nu_2)$, but only the flavor composition  of the lone state $\nu_3$, 
\begin{eqnarray}
\nu_3 	& = & 
	U_{e3}\,\nu_e +  U_{\mu3}\,\nu_\mu + U_{\tau3}\,\nu_\tau	\\
   	& = & 
   	s_\phi\,\nu_e + c_\phi (s_\psi\,\nu_\mu + c_\psi\,\nu_\tau)\ .   
\label{nu3}
\end{eqnarray}
The neutrino oscillation probabilities in vacuum assume then the
simple form:
\begin{eqnarray}
P^{\rm vac}({\nu_\alpha\leftrightarrow\nu_\alpha}) 
	& = & 
1-4U^2_{\alpha3}(1-U^2_{\alpha3})\,S\ ,\label{P1}
	\\
P^{\rm vac}({\nu_\alpha\leftrightarrow\nu_\beta}) 
	& = & 
4 U^2_{\alpha 3}U^2_{\beta 3}\, S\ \ (\alpha\neq\beta)
	\ ,
\label{P}
\end{eqnarray}
where
\begin{equation}
	S=
\sin^2\left(1.27 \frac {m^2\,[{\rm eV}^2]\cdot L\,[{\rm km}]}
{E_\nu\, [{\rm GeV}]}\right)\ ,
\label{S}
\end{equation}
neutrino and antineutrino probabilities being equal. However, in matter  
$P(\nu)\neq P(\overline\nu)$, and the probabilities must be calculated
numerically for the Earth density profile. For a constant density,
they can still be calculated analytically \cite{3ATM}.

One can make contact with the familiar two-flavor oscillation  scenarios in
three limiting cases:
\begin{eqnarray}
\phi = 0 
		&\ \ \ \Rightarrow \ \ \ & 
\nu_3 = s_\psi\, \nu_\mu + c_\psi\,\nu_\tau\ \ 
({\rm pure\ }\nu_\mu\leftrightarrow\nu_\tau{\rm\ osc.})\ ,
\\
\psi = {\frac{\pi}{2}} 
		&\ \ \ \Rightarrow \ \ \ & 
\nu_3 = s_\phi\,\nu_e + c_\phi\,\nu_\mu\ \ 
({\rm pure\ }\nu_\mu\leftrightarrow\nu_e{\rm\ osc.})\ ,
\\
\psi = 0 	
		&\ \ \ \Rightarrow \ \ \ & 
\nu_3 = s_\phi\, \nu_e + c_\phi\,\nu_\tau\ \ 
({\rm pure\ }\nu_e\leftrightarrow\nu_\tau{\rm\ osc.})\ ,
\end{eqnarray}
with the corresponding, further identifications (valid only in the $2\nu$ 
cases): 
\begin{eqnarray}
\sin^2 2\theta_{\mu\tau} 	&\equiv& 4 s^2_\psi c^2_\psi\ ,\\
\sin^2 2\theta_{e\mu} 		&\equiv& 4 s^2_\phi c^2_\phi\ ,\\
\sin^2 2\theta_{e\tau} 		&\equiv& 4 s^2_\phi c^2_\phi\ .
\end{eqnarray}

Finally, we remind that for pure $\nu_\mu\leftrightarrow\nu_\tau$ 
oscillations the physics is symmetric under the replacement
$\psi\to\pi/2-\psi$, due to the absence of matter effects. Such effects
instead break the (vacuum) symmetry $\phi\to\pi/2-\phi$ for pure
$\nu_\mu\leftrightarrow\nu_e$ oscillations, for which the cases $\phi<\pi/4$
and $\phi>\pi/4$ are distinguishable.%
\footnote{Alternatively, one can fix $\phi <\pi/4$ and consider the cases
	$m^2>0$ or $m^2<0$ \protect\cite{Gonz}.}
In generic three-flavor cases, no specific symmetry exists in the presence of
matter and the mixing angles $\phi$ and $\psi$ must be taken in their full
range $[0,\,\pi/2]$. A full account of the symmetry properties of the
oscillation probability in $2\nu$ and $3\nu$ cases, both in vacuum and in
matter, can be found in Appendix~C of \cite{3ATM}.

\subsection{The atmospheric $\nu$ parameter space}

As previously said, under the hypothesis (\ref{appr}) the parameter  space of
atmospheric $\nu$'s is spanned by $(m^2,U_{e3},U_{\mu3},U_{\tau3})$.  At any
fixed value of $m^2$, the unitarity condition
\begin{equation}
U^2_{e3}+U^2_{\mu3}+U^2_{\tau3}=1
\label{unit}
\end{equation}
can be embedded in a triangle graph \cite{AcRe,3MSW,Lisi}, whose corners 
represent the flavor eigenstates, while a generic point inside the triangle 
represents the ``lone'' mass eigenstate $\nu_3$. By identifying the heights 
projected from $\nu_3$ with the square matrix elements $U^2_{e3}$, 
$U^2_{\mu3}$, and $U^2_{\tau3}$, the unitarity condition (\ref{unit}) is
automatically satisfied for a unit height triangle \cite{AcRe}.

Fig.~4 shows the triangle graph as charted by the coordinates 
$U^2_{\alpha3}$ (upper panel) or $(\phi,\psi)$ (lower panel). When $\nu_3$ 
(the mass eigenstate) coincides  with one of the corners (the flavor 
eigenstates),  the no oscillation case is recovered.  The sides correspond to
pure $2\nu$ oscillations. Inner points in the  triangle represent genuine
$3\nu$ oscillations.

\section{Two-flavor analysis}

In this Section we study first how the theoretical zenith  distributions are
distorted, in the presence of two flavor oscillations, with respect to the
``flat'' expectations  of Fig.~3. This introductory  study does not involve
numerical fits to the data, and helps to understand  which  features of
either $\nu_\mu\leftrightarrow\nu_\tau$ or $\nu_\mu\leftrightarrow\nu_e$
oscillations may be responsible for the observed SK zenith distributions. We
then fit the most recent SK data  (33 kTy) using a $\chi^2$ statistics, and
discuss the results.

\begin{figure}[t]
\begin{center}
\epsfig{bbllx=2.0truecm,bblly=3.2truecm,bburx=20truecm,bbury=26.5truecm,clip=,%
width=9truecm,file=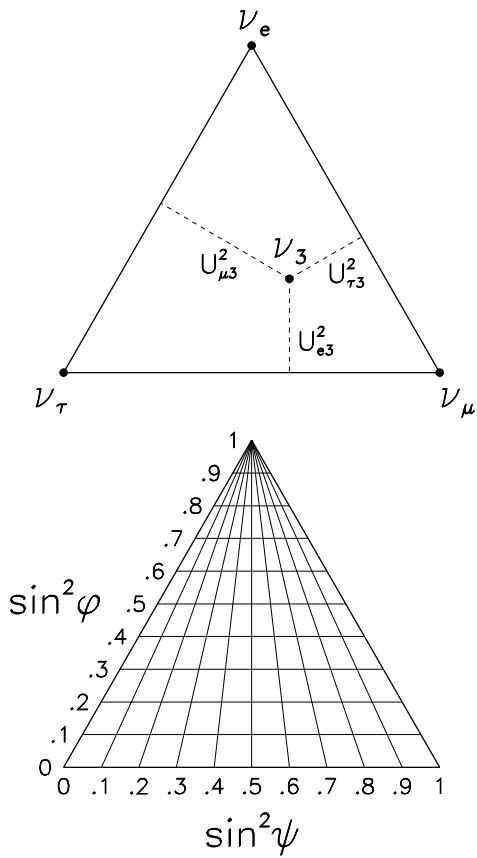}
\caption{Triangle graph representing the three-flavor mixing space of
$\nu_3$. In the upper panel the parameter space is
spanned by the matrix elements $U^2_{e3}$, $U^2_{\mu3}$, and
$U^2_{\tau3}$.  When such elements are identified with the heights projected
by a generic  point ($\nu_3$), the unitarity condition $\sum_\alpha
U^2_{\alpha3}=1$ is automatically satisfied (for a triangle of unit height).
Vertices, sides, and inner area correspond to no oscillation, two-flavor
oscillations, and three-flavor oscillations, respectively. In the lower panel,
the same parameter space is
charted through the mixing angles $\phi=\theta_{13}$ and
$\psi=\theta_{23}$.}
\end{center}
\end{figure}

\subsection{Zenith distributions for $\nu_\mu\leftrightarrow\nu_\tau$
oscillations}

Figure $5$ shows, in the same format as in Fig.~3,
our calculations of the five zenith distributions of
atmospheric neutrino events. In this and
in the following figures, the upper left box  contains comments on the
scenario, while the upper right box displays the selected values of
$(m^2/{\rm eV}^2,U^2_{e3},U^2_{\mu 3},U^2_{\tau 3})$. In Fig.~5 we 
consider, in particular, pure $\nu_\mu\leftrightarrow\nu_\tau$ oscillations 
$(U^2_{e3}=0)$ with maximal mixing  ($\sin^2
2\theta_{\mu\tau}=4U^2_{\mu3}U^2_{\tau 3}=1$). Of course, the electron
distributions are not affected by  $\nu_\mu\leftrightarrow\nu_\tau$
transitions, while the muon event rates are suppressed, especially for zenith
angles approaching the vertical ($\cos\theta=-1$, upward leptons),
corresponding to longer average  neutrino pathlengths. The prediction for
$m^2=10^{-3}$ eV$^2$ (dashed line)  is in reasonable agreement with all the
muon data samples (SG, MG, and UP).  For $m^2=10^{-2}$ eV$^2$ the expected
rates of SG$\mu$ and UP$\mu$ are  significantly suppressed,  and for
$m^2=10^{-1}$ eV$^2$ one approaches the  limit of energy-averaged $2\nu$
oscillations, with a flat suppression of  $\sim 50\%$, which does not appear in
agreement with the data. On the other hand,  decreasing $m^2$ down to
$10^{-4}$ eV$^2$ (thick, solid line), one has almost ``unoscillated''
distributions  for the high energy samples  MG$\mu$ and UP$\mu$, since the
phase $m^2L/E_\nu$ is small. At lower energies (i.e., SG$\mu$ events), 
however, this phase can still be large enough to  distort the zenith
distribution.

Notice that in Fig.~5 the theoretical electron distributions SG$e$ and MG$e$  
are always below the data points. Using the overall $\pm 30\%$
normalization freedom, one can imagine to ``rescale up'' all the five
theoretical distributions   (by, say, 15--20\%) to match the
electron data. This  upward shift would  also alter the muon distributions at
the same time, but one can easily  realize that such distributions  would
still be in reasonable agreement  with the data for $m^2=10^{-3}$ and
$10^{-2}$ eV$^2$. Therefore, at  $\sin^22\theta_{\mu\tau}=1$  one expects an
allowed range of $m^2$ around $m^2=10^{-3}$--$10^{-2}$ eV$^2$, independently
of the details of the statistical analysis. Values of $m^2$ outside this
range do not agree with the muon data.

\begin{figure}[t]
\begin{center}
\epsfig{bbllx=2.0truecm,bblly=9.4truecm,bburx=20truecm,bbury=19.4truecm,clip=,%
width=18truecm,file=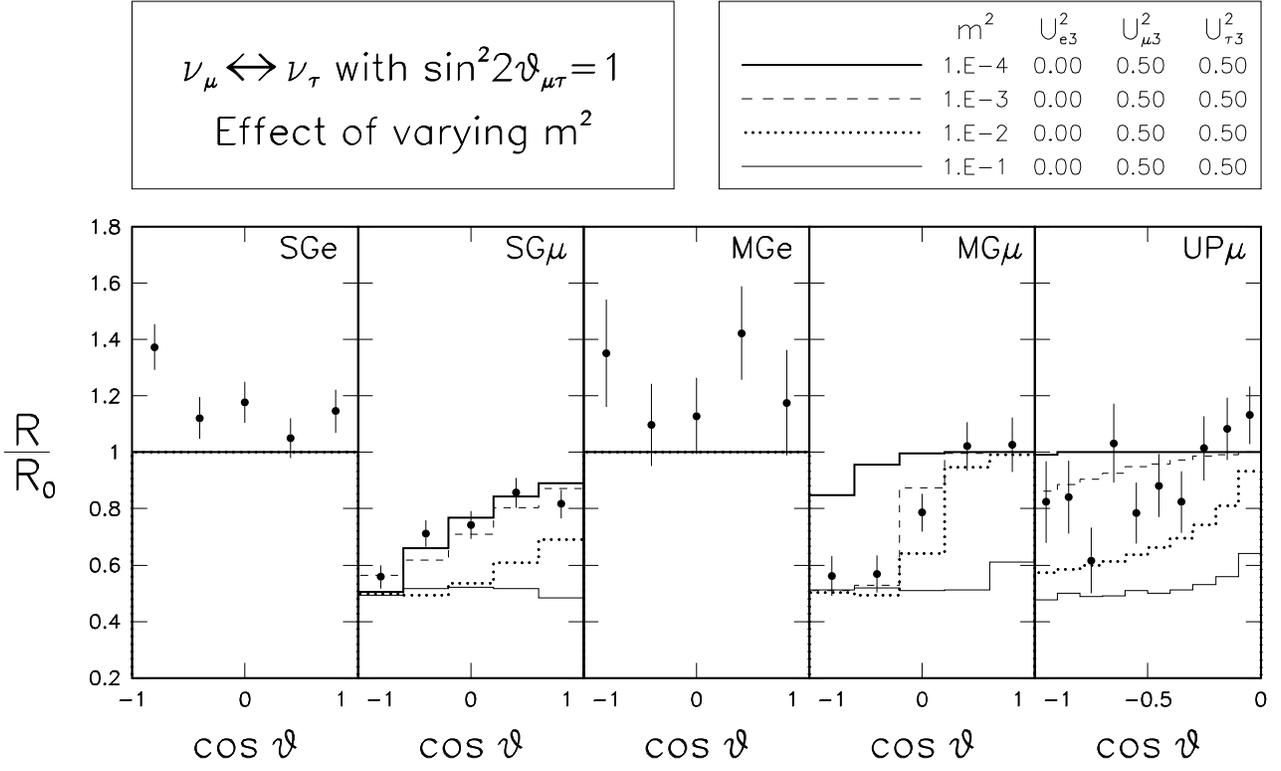}
\caption{Distortions of the zenith distributions induced by variations of
$m^2$ (eV$^2$), for pure $\nu_\mu\leftrightarrow\nu_\tau$ oscillations with
maximal $(\nu_\mu,\nu_\tau)$ mixing.}
\end{center}
\end{figure}
\begin{figure}[b]
\begin{center}
\epsfig{bbllx=2.0truecm,bblly=9.4truecm,bburx=20truecm,bbury=19.4truecm,clip=,%
width=18truecm,file=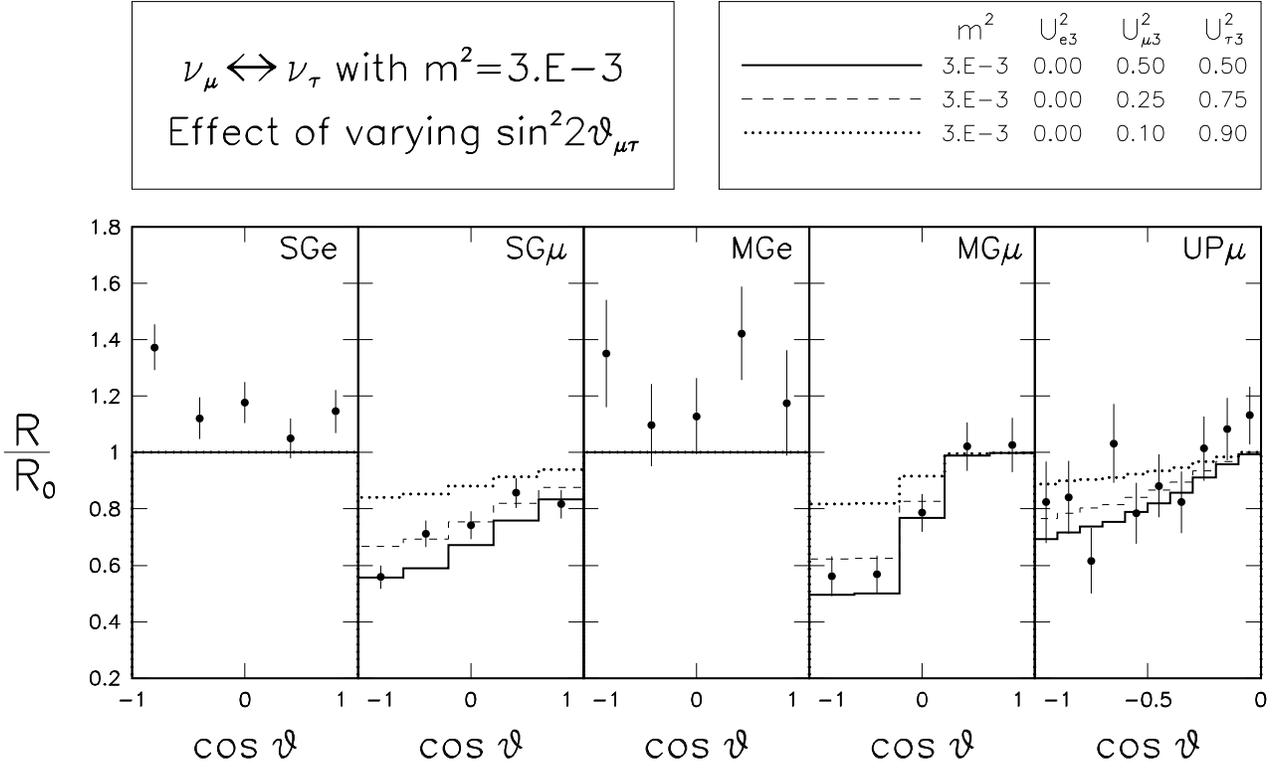}
\caption{Distortions of the zenith distributions induced by variations of
the $(\nu_\mu,\nu_\tau)$ mixing, for pure $\nu_\mu\leftrightarrow\nu_\tau$
oscillations with $m^2=3\times 10^{-3}$ eV$^2$.}
\end{center}
\end{figure}

As a final comment to Fig.~5, we notice that the theoretical rate for
downgoing multi-GeV muons (rightmost bin of the MG$\mu$ sample) is
practically identical to the unoscillated case for $m^2\lesssim 10^{-2}$ eV$^2$,
as also observed at the end of  Sec.~II~B. Therefore, the SK data can
potentially self-calibrate the {\em absolute\/} muon rate normalization {\em
independently of oscillations\/}, provided that the  total experimental 
error in the last MG$\mu$ bin is reduced to a few percent.

In Fig.~6 we take $m^2$ fixed (at $3\times10^{-3}$ eV$^2$) and  vary the
$\nu_\mu$-$\nu_\tau$ mixing ($\sin^2 2\theta_{\mu\tau}=4U^2_{\mu3}U^2_{\tau
3}=1,0.75,0.36$). The suppression of the muon rates increases with increasing
mixing; however, there is no dramatic difference between
$\sin^22\theta_{\mu\tau}=1$ and 0.75 (solid and dashed lines, respectively).
A value as low as $\sin^22\theta_{\mu\tau}=0.36$ is not in agreement with
the  SG and MG muon distributions, although it is still allowed by the
UP$\mu$ sample (which does not place strong bounds on the mixing).  If all
the distributions were renormalized (scaled up) to match the SG$e$ and MG$e$
samples, small mixing values would be even more disfavored. Therefore, we
expect the mixing angle to be in the range $\sin^2 2\theta_{\mu\tau}=0.8$--1,
independently on the details of the statistical analysis.

We emphasize that, although in principle $e$-like events are not affected
by $\nu_\mu\leftrightarrow\nu_\tau$ oscillations, the SG$e$ and MG$e$ 
samples affect indirectly the estimate of the mass-mixing  parameters, since
they drive the fit to higher values of the neutrino fluxes.  Further
experimental constraints on the overall neutrino flux normalization might
have then a significant impact on the current estimates of  $m^2$ and $\sin^2
2\theta_{\mu\tau}$.

\vspace*{-2mm}
\subsection{Zenith distributions for $\nu_\mu\leftrightarrow\nu_e$
oscillations}

Fig.~7 is analogous to Fig.~5, but for $\nu_\mu\leftrightarrow\nu_e$
oscillations with maximal mixing ($U^2_{\tau3}=0$  and
$\sin^2\theta_{e\mu}=4U^2_{e3}U^2_{\mu3}=1$). Matter effects are included.
The expected SG and MG rates of electrons coming from below  appear to be
enhanced, due to $\nu_\mu$'s oscillating into  $\nu_e$'s. The slope of the
zenith  distribution is stronger for MG$e$ than  for SG$e$, because:
(i) 	The $\nu_\mu/\nu_e$ flux ratio  increases with energy, as observed in
Section~II~A; and  (ii) 	the ``angular smearing'' due to the different
lepton and neutrino directions is more effective for SG events.  On the other
hand, the suppression of the muon rates is not as effective as for the
$\nu_\mu\leftrightarrow\nu_\tau$ case in Fig.~5. In fact, now there  are some
$\nu_e$'s oscillating back into $\nu_\mu$'s. Moreover,  matter effects tend
to {\em suppress\/} large-amplitude  oscillations (when the mixing is maximal
in vacuum, it can  only be smaller in matter). In general, one has a too
strong increase of  electrons and a too weak suppression of muons, although
this pattern may be in part improved by {\em rescaling down\/} all the
theoretical curves.  In this case, the distributions at $m^2=10^{-3}$ and
$10^{-2}$ eV$^2$ can  get in marginal agreement with all the data, while
$m^2=10^{-4}$ eV$^2$ is in any case excluded. Notice, however,  that
$m^2=10^{-2}$ eV$^2$ is not  allowed by CHOOZ  \cite{Cho1}.

\begin{figure}[b]
\begin{center}
\epsfig{bbllx=2.0truecm,bblly=9.5truecm,bburx=20truecm,bbury=19.4truecm,clip=,%
width=18truecm,file=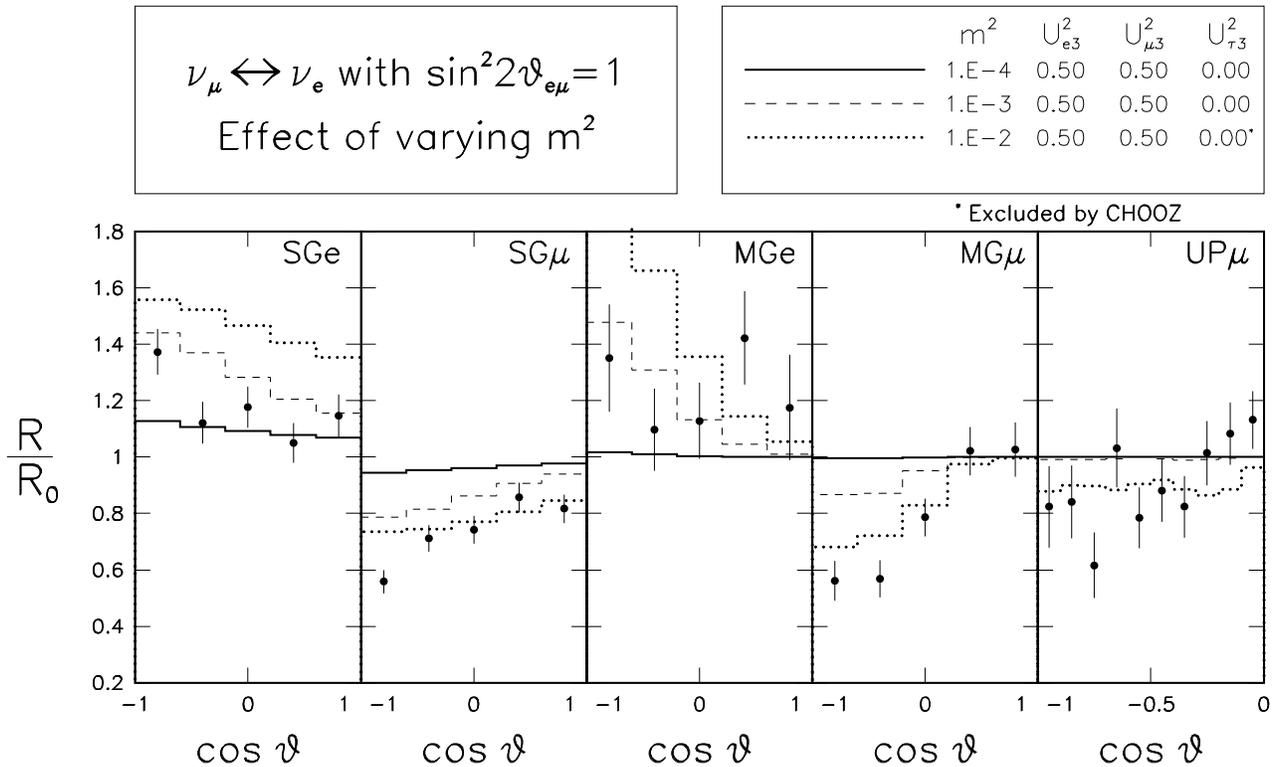}
\caption{Distortions of the zenith distributions induced by
variations of $m^2$ (eV$^2$), for pure $\nu_\mu\leftrightarrow\nu_e$
oscillations with maximal $(\nu_\mu,\nu_e)$ mixing.}
\end{center}
\end{figure}

\begin{figure}[t]
\begin{center}
\epsfig{bbllx=2.0truecm,bblly=9.4truecm,bburx=20truecm,bbury=19.4truecm,clip=,%
width=18truecm,file=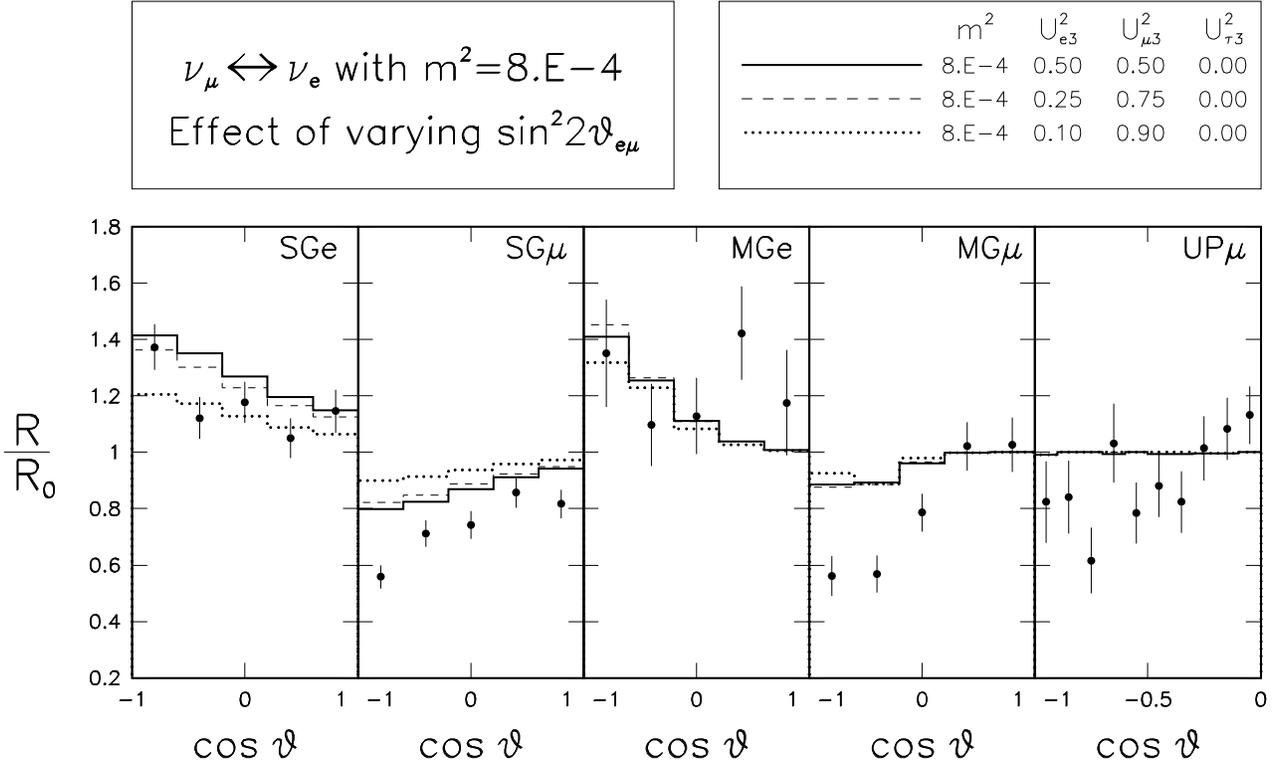}
\caption{Distortions of the zenith distributions induced by variations of
the $(\nu_\mu,\nu_e)$ mixing, for  pure $\nu_\mu\leftrightarrow\nu_e$
oscillations with $m^2=8\times 10^{-4}$ eV$^2$ (i.e., below CHOOZ bounds).}
\end{center}
\end{figure}
\begin{figure}[b]
\begin{center}
\epsfig{bbllx=2.0truecm,bblly=9.4truecm,bburx=20truecm,bbury=19.4truecm,clip=,%
width=18truecm,file=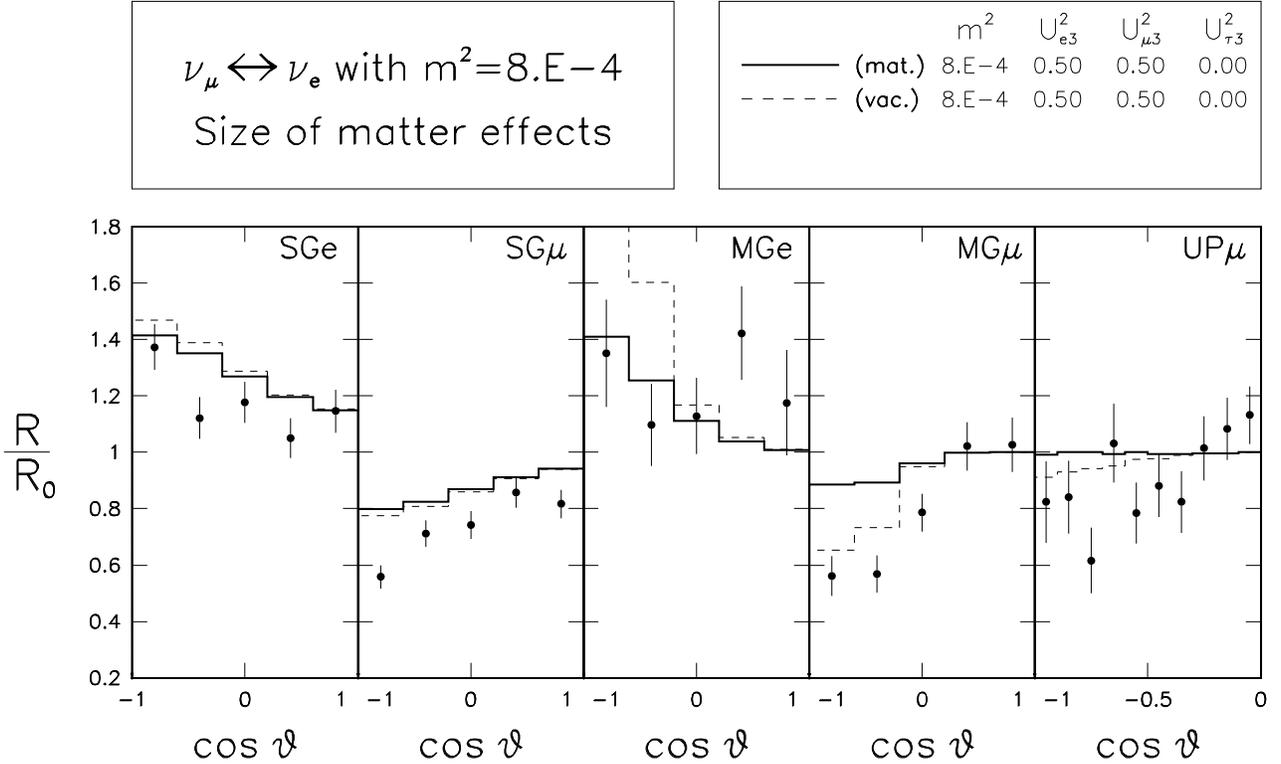}
\caption{Pure $\nu_\mu\leftrightarrow\nu_e$ oscillations with $m^2=8\times
10^{-4}$ eV$^2$ (i.e., below CHOOZ bounds) and maximal $(\nu_\mu,\nu_e)$
mixing, with and without matter effects (solid and dashed lines, respectively.}
\end{center}
\end{figure}

Figure~8 shows the effect of varying the $\nu_\mu\leftrightarrow\nu_e$
mixing, for $m^2=8\times 10^{-4}$ eV$^2$---a value safely below the CHOOZ 
bounds. It can be seen that variations of the mixing do not help much in 
reaching an agreement with the data for this value of $m^2$, which thus 
seems to be disfavored.

Figure~9 shows the size of matter effects for $m^2=8\times 10^{-4}$ eV$^2$ and
maximal $\nu_\mu\leftrightarrow\nu_e$ mixing. For values of $m^2$ around 
$10^{-3}$ eV$^2$ such as this,  matter effects are more important for the
multi-GeV sample, while  lower $m^2$ values would enhance the effect for the
sub-GeV sample and higher values for the UP$\mu$ sample (see, e.g., \cite{Marr}
and  Figs.~4 and 6 in \cite{3ATM}).  It can be seen that, in any case, the net
matter effect is a decrease of the slope of the zenith distribution for both
muons  and electrons, i.e., a suppression of the oscillation amplitude with
respect to the ``pure vacuum'' case. The effect is not completely reduced to
zero above horizon, due to the neutrino-lepton angular smearing.

We summarize the content of Figs.~7--9 by observing that the theoretical
zenith distributions, in the presence of  $\nu_\mu\leftrightarrow\nu_e$
oscillations, are at most in marginal  agreement with the SK data set.  The
agreement is somewhat improved by rescaling down the expectations. Earth
matter effects  are sizable and cannot be neglected in the analysis.

\subsection{Fits to the data}

In the previous two subsections we have presented quantitative calculations
of the zenith distributions in selected two-flavor scanarios, and a
qualitative comparison with the data. Here we discuss the results of a
quantitative $2\nu$ fit to the SK data.

Figure~10 shows the results of our $\chi^2$ analysis of the SK data (SG$e$,
SG$\mu$, MG$e$, MG$\mu$, and UP$\mu$ data combined).  The panel (a) refers to
$\nu_\mu\leftrightarrow\nu_\tau$ oscillations  ($\phi=0$) in  the plane
($m^2$,\,$\sin^2 2\psi$). We find a minimum value  $\chi^2_{\rm min}=29.6$
for 28 degrees of freedom (30 data points minus 2 oscillation parameters),
indicating a good fit to the data.%
\footnote{The best-fit value of $m^2$ is not very meaningful at present. 
	We prefer to focus on confidence intervals.}
This is to be contrasted to the value  $\chi^2_{\rm min}=126$ for the no 
oscillation case, which is  therefore excluded by the SK data with very  high
confidence. The quantitative limits on the mass-mixing parameters are 
consistent with the qualitative expectations discussed in Section~IV~A. 
Moreover, the allowed region is in agreement with the global 
analysis of pre-SK data shown in Fig.~2 of \cite{3ATM}.%
\footnote{In \protect\cite{3ATM} we obtained an allowed range for $m^2$ 
	larger than the one reported  by the Kamiokande Collaboration
	\protect\cite{Ymul}, presumably as a result of a different approach
	to the statistical analysis \protect\cite{3ATM,Stat}.
	See also the comments of \protect\cite{Salt} about the  
	Kamiokande bounds in \protect\cite{Ymul}.}
Our allowed range of $m^2$ in Fig.~10(a) is somewhat narrower than the  range
estimated by the SK Collaboration \cite{EVID}. We have checked that the
differences are largely due to the fact that only SG and MG were fitted in
\cite{EVID}, while here we include also UP$\mu$ data, which help to exclude
the lowest values of $m^2$  (see also \cite{Kaji}).  To a lesser extent,  our
different definition of $\chi^2$ (see Appendix~B) also plays a role.

Figures 10(b) and 10(c) refer to  $\nu_\mu\leftrightarrow\nu_e$  oscillations
($\psi=\pi/2$) in the plane ($m^2$,\,$\sin^2 2\phi$), for  $\phi<\pi/4$ and
$\phi>\pi/4$, respectively (the two cases being different,  see Sec.~III~B).
The minimum value of $\chi^2$ is now much higher (67.7 and 68.6 for panel (b)
and (c), respectively) indicating that $\nu_\mu\leftrightarrow\nu_e$
oscillations are disfavored by the SK data. This represents an important step
forward with repect to 
pre-SK data, which did not distinguish significantly between
$\nu_\mu\leftrightarrow\nu_\tau$ and   $\nu_\mu\leftrightarrow\nu_e$
oscillations \cite{3ATM} (and, actually, showed a slight preference for the
latter \cite{3ATM,Marr}). In Figs.~10(b,c), the C.L.\ limits around the
minimum appear to be shifted to higher values of $m^2$ if compared to
Fig.~10(a),  as a consequence of matter effects that suppress the effective
mixing in the lowest range of $m^2$ (see also \cite{3ATM}).

The $\nu_\mu\leftrightarrow\nu_e$ allowed regions in Fig.~10(b) and  10(c)
have a relatively scarce interest. On the one hand, they represent a  poor
fit to the SK data themselves. On the other hand, they are excluded by the
CHOOZ reactor experiment. Figure~10(d) shows the CHOOZ bounds, as  derived by
our own $\chi^2$ reanalysis, in good agreement with the limits  shown in
\cite{Cho2}. Such bounds are in contradiction with the 
$\nu_\mu\leftrightarrow\nu_e$ allowed regions in Fig.~10(b,c) at 90\% C.L.,
(although there might be a marginal agreement at 99\% C.L.); therefore, we do
not make any attempt to combine SK+CHOOZ data in a $2\nu$ analysis.

Summarizing, our results for two-flavor oscillations are consistent 
with the $2\nu$ analysis of the SK Collaboration \cite{EVID}, namely: 
(i) 	The no oscillation hypothesis is rejected with high confidence; and 
(ii)	$\nu_\mu\leftrightarrow\nu_\tau$ oscillations are
	largely preferred over $\nu_\mu\leftrightarrow\nu_e$. 
On our part, we add the following nontrivial statement: 
(iii) 	The present SK bounds on the  $\nu_\mu\leftrightarrow\nu_\tau$ 
	mass-mixing parameters are in good agreement with those obtained 
	from the global analysis of pre-SK data in \cite{3ATM} (including 
	NUSEX, Fr\'ejus, IMB, and Kamiokande sub-GeV and multi-GeV data).

\begin{figure}[b]
\begin{center}
\epsfig{bbllx=2.0truecm,bblly=6truecm,bburx=20truecm,bbury=25truecm,clip=,%
width=18truecm,file=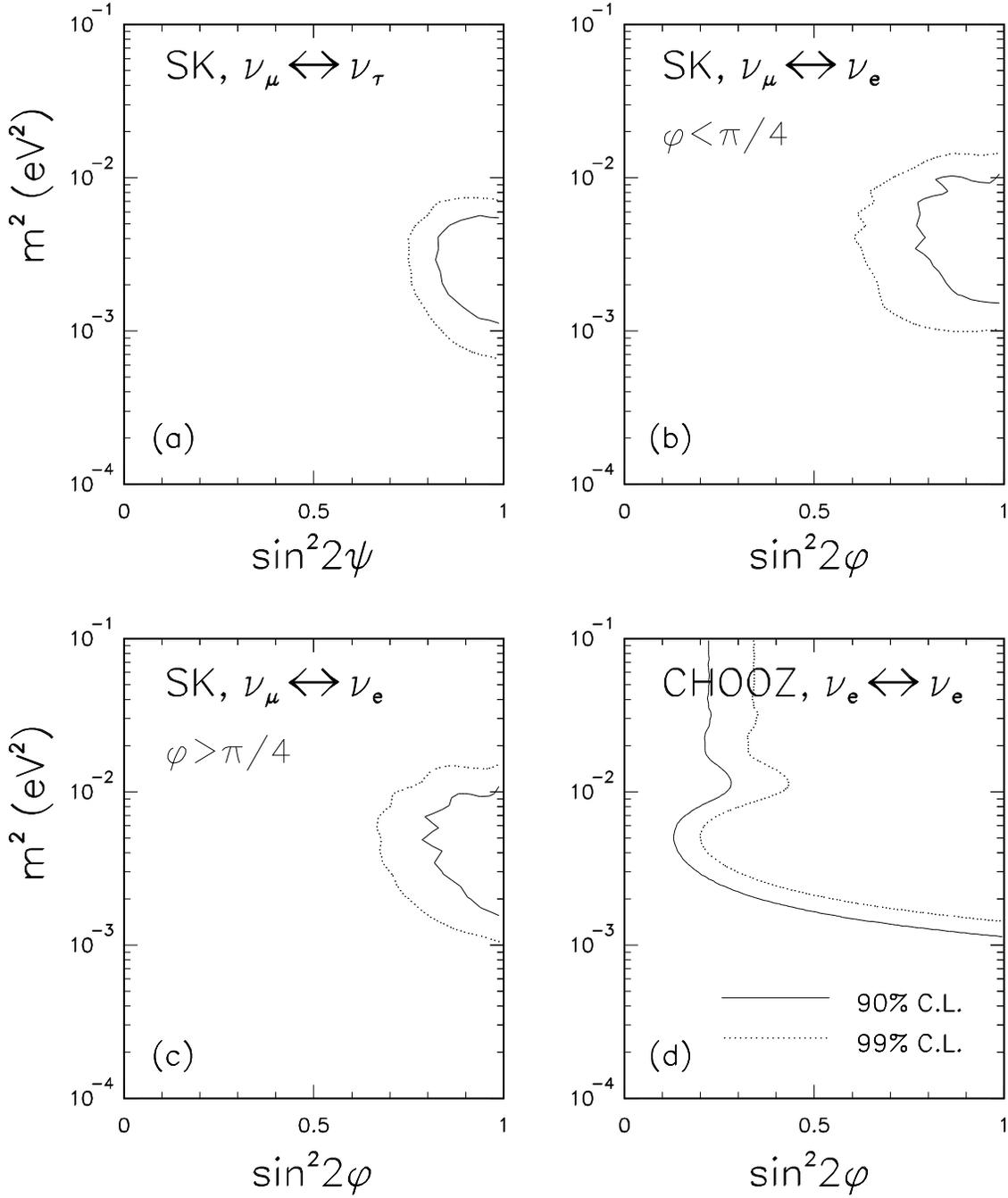}
\caption{Two-flavor oscillation fits to the SK 
zenith distributions (SG, MG, and UP$\mu$ combined). (a) Fit for
$\nu_\mu\leftrightarrow\nu_\tau$ ($\phi=0$) in the plane $(m^2,\sin^22\psi)$.
The cases $\psi<\pi/4$ and  $\psi>\pi/4$ are equivalent. (b) Fit for 
$\nu_\mu\leftrightarrow\nu_e$  ($\psi=\pi/2$) in the plane
$(m^2,\sin^22\phi)$, for $\phi<\pi/4$. (c) Fit for 
$\nu_\mu\leftrightarrow\nu_e$  ($\psi=\pi/2$) in the plane
$(m^2,\sin^22\phi)$, for $\phi>\pi/4$. The cases (b) and (c) are
different, due to earth  matter effects. The limits coming from the CHOOZ
experiment are also shown  in panel (d), as derived by our reanalysis. The
solid and dotted curves correspond to 90\% and 99\% C.L., i.e., to variations
of $\chi^2-\chi^2_{\rm min}= 4.61,\,9.21$ for two degrees of freedom (the
oscillation parameters).}
\end{center}
\end{figure}

\bigskip
\bigskip
\bigskip
\bigskip

Finally, we comment upon recent claims \cite{Lose} of  inconsistencies within
the SK data, under the hypothesis of  $\nu_\mu\leftrightarrow\nu_\tau$
oscillations. The argument goes as follows.  If 
$\nu_\mu\leftrightarrow\nu_\tau$ oscillations are assumed,  the electron
excess has to be adjusted by rescaling up the theoretical predictions. Then
the overall muon deficit becomes close to  maximal ($\sim 50\%$) for the SG
and MG data (see our Fig.~2, middle panel).  Such large suppression of the
muon rate seems to suggest energy-averaged  oscillations (i.e., large $m^2$).
On the other hand, the zenith distortions  {\em require\/} energy-dependent
oscillations (i.e., relatively small $m^2$). Some difference between the
``total rate'' and the ``shape'' information,  although exacerbated by the
semiquantitative calculations of \cite{Lose}, indeed exists but is not
entirely new, as it has already been investigated by the SK Collaboration
\cite{Kaji}. In fact, the slide No.~18 of \cite{Kaji} shows the separate
$\nu_\mu\leftrightarrow\nu_\tau$ fits to the ``total rate''  and ``shape''
data, the former preferring values of $m^2$ tipically higher than the latter.
However, in the same slide \cite{Kaji} one can see a  reassuring, large 
overlap between the two allowed ranges  (at $90\%$ C.L.) for $m^2\sim {\rm
few}\times 10^{-3}$ eV$^2$. This means that, within the present (relatively
large) experimental and  theoretical uncertainties, there is no real
contradiction between different pieces of SK data, as also confirmed by our
good $\nu_\mu\leftrightarrow\nu_\tau$ global fit to the SK data. However, 
the above remarks,  as well as our comments on the absolute event rates in
Sec.~II, should be kept in mind when new, more accurate experimental or
theoretical information will become available.

\section{Three-flavor analysis}

In this Section we discuss in detail a three-flavor analysis of the SK and
CHOOZ data. In the first subsection we show  representative  examples of $3\nu$
oscillation  effects on the zenith distributions. In the second  subsection we
discuss some issues related to the $L/E_\nu$ variable. In the  third and fourth
subsections we report the results of detailed fits to the  SK data, without and
with the additional constraints from the CHOOZ  experiment, respectively.

We remind that, in three flavors, the CHOOZ mixing parameter
$\sin^22\theta_{ee}$ can be identified with $4U^2_{e3}(1-U^2_{e3})$
[Eq.~(\ref{P1})]; therefore, the CHOOZ constraint  $\sin^2
2\theta_{ee}\lesssim 0.22$, valid  for $m^2\gtrsim 2\times 10^{-3}$ eV$^2$
\cite{Cho2}, translates into either $U^2_{e3}\lesssim 0.06$ or 
$U^2_{e3}\gtrsim 0.94$.

\subsection{Zenith distributions: Expectations for three-flavor  oscillations}

In a three-flavor language, two-flavor oscillations with maximal 
$\nu_\mu\leftrightarrow\nu_\tau$  mixing are characterized by 
$(U^2_{e3},U^2_{\mu3},U^2_{\tau3}=0,1/2,1/2)$ (the center of the lower side 
in the triangle graph of Fig.~4). Analogously, maximal 
$\nu_\mu\leftrightarrow\nu_e$ mixing  is characterized by 
$(U^2_{e3},U^2_{\mu3},U^2_{\tau3}=1/2,1/2,0)$ (the center of the right side
of the triangle). A smooth $3\nu$ interpolation  between these $2\nu$ cases
can be performed by gradually increasing the  value of $U^2_{e3}$ from 0  to
1/2,  and decreasing the value of  $U^2_{\tau3}$ from 1/2 to 0 at the same
time, the element $U^2_{\mu3}$ being adjusted to preserve unitarity.

This exercise is performed in Fig.~11 for a relatively high value of $m^2$
($m^2=8\times 10^{-2}$ eV$^2$). The thick and thin solid  lines represent the
zenith distributions for pure  $\nu_\mu\leftrightarrow\nu_\tau$ and pure 
$\nu_\mu\leftrightarrow\nu_e$  oscillations with maximal mixing,
respectively. The dashed and dotted lines  represent intermediate $3\nu$
cases, the first being ``close'' to  $\nu_\mu\leftrightarrow\nu_\tau$ (with
an additional $20\%$ admixture of  $\nu_e$), and the second being ``close''
to $\nu_\mu\leftrightarrow\nu_e$ (with an additional $20\%$ admixture of
$\nu_\tau$). Although none of the four cases depicted in Fig.~11 represents a
good fit  to all the SK data, and three of them are excluded by  CHOOZ,  much
can be learned from a qualitative understanding of the zenith  distributions
in this figure.

In the presence of oscillations,  the distributions $R/R_0$ in  Fig.~11 are
roughly given by 
\begin{eqnarray}
\frac{\mu}{\mu_0} & \sim & P_{\mu\mu} + \frac{e_0}{\mu_0}\,P_{e\mu}\ ,
\label{app1}\\
\frac{e}{e_0} & \sim & P_{ee} + \frac{\mu_0}{e_0}\,P_{\mu e}\ ,\label{app2}
\end{eqnarray}
$\mu_0$ and $e_0$ being the unoscillated rates. For a relatively high value of
$m^2$ as that in Fig.~11, the asymptotic regime  of energy-averaged oscillations
approximately  applies, except for the rightmost bins of the MG and UP$\mu$
distributions  (where the shorter pathlengths require higher $m^2$'s for
reaching such  regime). Then, for pure $\nu_\mu\leftrightarrow\nu_\tau$
oscillations with maximal mixing,  one  has $P_{ee}=1$, $P_{e\mu}=0$, and
$P_{\mu\mu}\sim 1/2$, so that $e/e_0=1$  and $\mu/\mu_0\sim 1/2$, as indicated
by the thick, solid lines in Fig.~11.

For pure $\nu_\mu\leftrightarrow\nu_e$ oscillations with  maximal mixing, one
has $P_{ee}\sim 1/2$,  $P_{e\mu}\sim 1/2$, and $P_{\mu\mu}\sim 1/2$, so that 
$\mu/\mu_0\sim(1+e_0/\mu_0)/2$ and $e/e_0\sim (1+\mu_0/e_0)/2$. For sub-GeV
data, the often-quoted value $\mu_0/e_0\sim 2$ applies, so that $e/e_0\sim 1.5$
and $\mu/\mu_0\sim 0.75$, as indicated by the thin, solid lines in the SG$e$
and SG$\mu$ panels of Fig.~11. For multi-GeV data, however, the value
$\mu_0/e_0$ is not constant, ranging from $\sim 2$ along the horizontal
$(\cos\theta=0)$ to $\sim 3$ along the vertical $(\cos\theta=\pm 1$, see also
Fig.~1). Therefore, the ratio $\mu/\mu_0$ decreases slightly from $\sim 0.75$ 
(horizontal) to $\sim 0.67$ (vertical), while the ratio $e/e_0$ increases 
significantly from $\sim 1.5$ (horizontal) to $\sim 2$ (vertical). This is 
particularly evident as a ``convexity'' of the MG$e$ distribution in Fig.~11 
(thin, solid line).  Such behavior is not confined to pure two-flavor
oscillations, but is also present in genuine three-flavor cases, as indicated
by the dashed and dotted lines in the MG$e$ panel of Fig.~11. This shows that,
in the presence of $(\nu_\mu,\nu_e)$ mixing ($U^2_{e3}\neq0$), the variations
of the unoscillated ratio $\mu_0/e_0$  with $\cos\theta$ induce distortions of
the zenith distributions even in  the regime of energy-averaged oscillations
\cite{Zeni}, contrary to naive expectations. Notice that such distortions do
not depend on $L/E_\nu$, but on $L$ and $E_\nu$ separately through
$\mu_0/e_0=\mu_0/e_0(L(\theta),E_\nu)$.

Another peculiar distortion, not dependent on $L/E_\nu$, is related to a
genuine three-flavor effect in matter \cite{Zeni,Giun}. This  effect is
basically due to the splitting of the quasi-degenerate doublet 
$(\nu_1,\nu_2)$ in matter which, in the limit of large $m^2$, leads to an 
effective square mass difference  $\delta m^2_{\rm mat} \propto E_\nu$ and 
thus to a subleading  oscillation phase  $\delta m^2_{\rm mat}L/E_\nu\propto
L$ which  does not depend on $E_\nu$ but  only on $L$ \cite{Zeni}. The main
effect, relevant for atmospheric neutrinos, is to {\em decrease\/} the
$\nu_\mu$ survival probability $P_{\mu\mu}$ by an amount $\delta P$ which,
for a constant electron density $N_e$, reads  \cite{Zeni}
\begin{equation}
\delta P = 4 \frac{U^2_{e3}U^2_{\mu3}U^2_{\tau 3}}{(1-U^2_{e3})^2}
\sin^2\left(2.47(1-U^2_{e3}) \frac{N_e}{{\rm mol/cm}^3}\cos\Theta\right)\ .
\label{dP}
\end{equation}
Notice that the oscillation amplitude can be sizable only for large
three-flavor mixing, while it disappears for two-flavor mixing (i.e., when
one of the $U^2_{\alpha3}$  is zero), as indicated by the comparison of the
dotted and thin solid lines in the MG$\mu$ and UP$\mu$ panels of Fig.~11. 
The phase of $\delta P$ (the argument of $\sin^2$) can be rather large in the
Earth matter  ($N_e\simeq 2$--6 mol/cm$^3$) and is modulated by the neutrino
zenith angle  $\Theta$.  This modulation is particularly evident in the 
genuine $3\nu$ cases of the UP$\mu$ panel in Fig.~11  (dotted and dashed
lines), since upward through-going muons are highly correlated in direction
with the parent neutrinos  ($\theta\simeq\Theta$).  The $\cos\Theta$
modulation of $\delta P$ is  increasingly smeared out in the lower energy MG
and SG muon samples.

\bigskip
\bigskip
\bigskip
\bigskip

\begin{figure}[b]
\begin{center}
\epsfig{bbllx=2.0truecm,bblly=9.4truecm,bburx=20truecm,bbury=19.4truecm,clip=,%
width=18truecm,file=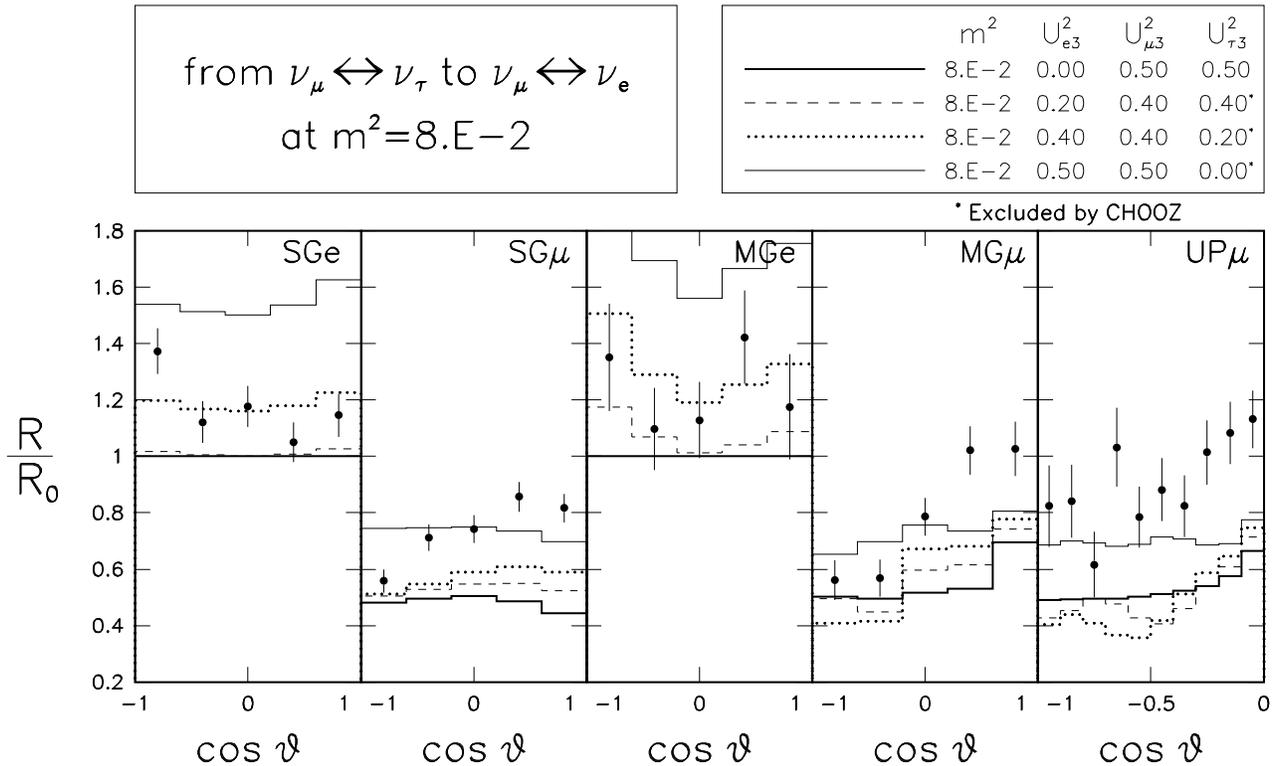}
\caption{Three-flavor oscillations at $m^2=8\times 10^{-2}$ eV$^2$. 
Distortions of the zenith distributions induced by variations of the 3$\nu$
mixing. The values of $U^2_{\alpha3}$ interpolate smoothly between the pure
$2\nu$ subcases. Some cases are excluded by CHOOZ.}
\end{center}
\end{figure}

We remind that in Fig.~11 the dashed curves ($U^2_{e3}=0.2$) correspond to a
$3\nu$ ``perturbation'' of pure  $\nu_\mu\leftrightarrow\nu_\tau$
oscillations (thick solid curves, $U^2_{e3}=0$), while the dotted curves
($U^2_{\tau3}=0.2$) correspond to a $3\nu$ ``perturbation'' of pure 
$\nu_\mu\leftrightarrow\nu_e$ oscillations (thin solid curves,
$U^2_{\tau3}=0)$. By comparing the dashed and thick  solid curves, it can be
seen that $3\nu$ perturbations of  $\nu_\mu\leftrightarrow\nu_\tau$
oscillations do not alter dramatically the  zenith distributions. The
opposite happens for the  $\nu_\mu\leftrightarrow\nu_e$ case (dotted and thin
solid curves). This  pattern can be explained as an interference between
vacuum and matter  effects.

More precisely, let us consider the case of large $m^2$  and $\mu_0/e_0\sim
2$. A $3\nu$ perturbation of   $\nu_\mu\leftrightarrow\nu_\tau$ oscillations
with maximal mixing can be parametrized by taking  $U^2_{e3}=\epsilon$ and 
$U^2_{\mu3}=U^2_{\tau3}=(1-\epsilon)/2$ (the $2\nu$ case being recovered  for
$\epsilon = 0$). The relevant oscillation probabilities are then
$P_{ee}\sim 1-2\epsilon$, $P_{e\mu}\sim \epsilon$, and
$P_{\mu\mu}\sim1/2-\delta P$ [notice that $\delta P$ is of $O(\epsilon)$,
see Eq.~(\ref{dP})]. Equations~(\ref{app1},\ref{app2}) give the muon and 
electron rates, $\mu/\mu_0\sim 1/2+\epsilon/2-\delta P$ and
$e/e_0\sim 1$, respectively. Therefore,
to  the first order in $\epsilon=U^2_{e3}$, the
electron rate does not vary, and the muon rate varies little since
$\epsilon/2$ and $\delta P$ partly cancel.

Conversely, a $3\nu$ perturbation of  $\nu_\mu\leftrightarrow\nu_e$ 
oscillations with maximal mixing can be parametrized by taking   $U^2_{\tau
3}=\epsilon$ and $U^2_{e3}=U^2_{\mu3}=(1-\epsilon)/2$ (the  $2\nu$ case being
recovered for $\epsilon = 0$). The relevant oscillation probabilities are then
$P_{ee}\sim 1/2$, $P_{e\mu}\sim 1/2-\epsilon$, and $P_{\mu\mu}\sim1/2-\delta
P$. The  muon and electron rates are now given by $\mu/\mu_0\sim
3/4-\epsilon/2-\delta P$ and $e/e_0\sim 3/2-2\epsilon$, respectively.  To the
first order in  $\epsilon=U^2_{\tau 3}$, the electron rate decreases, and also 
the  muon rate is suppressed, since the  terms $\epsilon$ and $\delta P$ have 
the  same sign. Therefore, adding some $\nu_\tau$ mixing ($U^2_{\tau 3}\neq 0$)
to $\nu_\mu\leftrightarrow\nu_e$ oscillations changes the predictions
considerably (generally  in the direction of a better fit to the data). For
this reason, we expect significant changes in the fit to SK data when moving
continuously from pure $\nu_\mu\leftrightarrow\nu_e$  oscillations to genuine
$3\nu$ cases.

For simplicity, we have discussed the above three-flavor effects at  large
$m^2$. Values of $m^2$ lower than in Fig.~11  are more interesting
phenomenologically (being less constrained by CHOOZ) but more difficult to
understand qualitatively, since the oscillations are no longer
energy-averaged, and  $2\nu$, $3\nu$, vacuum, and matter  effects are
entangled. Numerical calculations are required, and the results for
$m^2=8\times 10^{-3}$ and  $8\times 10^{-4}$ eV$^2$ are shown in Figs.~12 and
13, respectively (with mixing values chosen as in Fig.~11). The value  
$m^2=8\times 10^{-4}$ eV$^2$ is safely below the CHOOZ bounds. In Figs.~12 
and 13, the genuine $3\nu$ cases (dashed and dotted curves) show an improved 
agreement with the data (with respect to pure
$\nu_\mu\leftrightarrow\nu_\tau$ oscillations), since they give an excess of
electrons without  perturbing too much the muon distributions. Although this
advantage is not decisive at present, in view of the large 
uncertainties affecting the absolute normalization of the lepton rates, 
it might become crucial when such uncertainties will be reduced.

\subsection{Is $L/E_\nu$ a good variable?}

If two-flavor $\nu_\mu\leftrightarrow\nu_\tau$ oscillations  $(U^2_{e 3}=0)$
were the true and exclusive explanation of the SK atmospheric data, then it 
would make sense to try to reconstruct the (unobservable)  $L/E_\nu$ 
distribution of parent neutrinos from the lepton energies and directions. In
fact, any distortion effect should be related to vacuum oscillations and thus
to the ratio $L/E_\nu$ rather  than to $L$ and $E_\nu$ separately. The
theoretical and experimental $L/E_\nu$ distributions for SK can be found  in
\cite{EVID}.

However, beyond the $2\nu$ approximation, there are several  oscillation
effects that do not depend on $L/E_\nu$. Some of these effects,  originating
from $\nu_e$ mixing $(U^2_{e 3} \neq 0)$,   have been described in the
previous subsection.  Non-$L/E_\nu$ effects also arise in the presence of two
comparable square mass differences (i.e., $\delta m^2\sim m^2$ instead of 
$\delta m^2\ll m^2$), or if non-oscillatory phenomena contribute to 
partially explain the data.

Therefore, plots in the $L/E_\nu$ variable convey correct and unbiased
information only under the hypothesis of pure 
$\nu_\mu\leftrightarrow\nu_\tau$ oscillations. In all other cases, including 
our $3\nu$ framework, such plots cannot be used consistently. The following 
$3\nu$ fits, as for the $2\nu$ cases, make use of the zenith distributions 
and not of the reduced $L/E_\nu$ information.

\begin{figure}[b]
\begin{center}
\epsfig{bbllx=2.0truecm,bblly=9.4truecm,bburx=20truecm,bbury=19.4truecm,clip=,%
width=18truecm,file=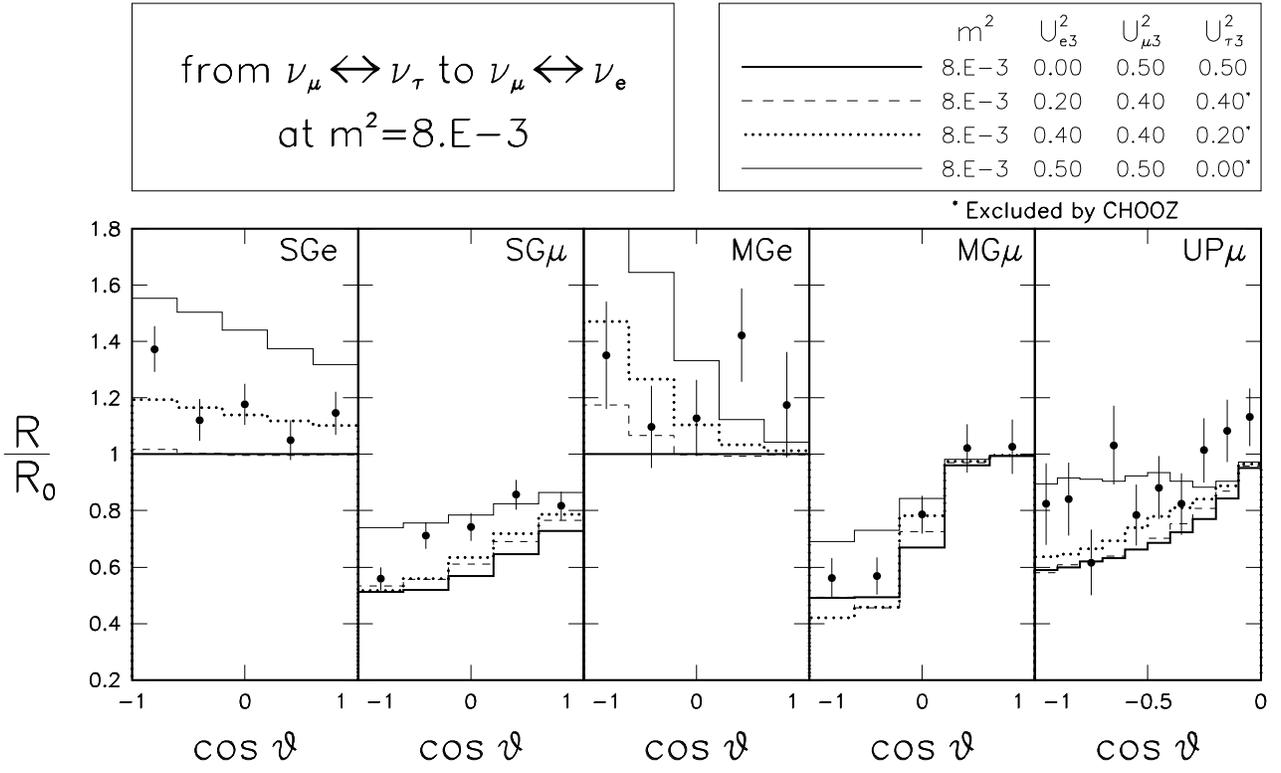}
\caption{As in Fig.~11, but for $m^2=8\times 10^{-3}$ eV$^2$.}
\end{center}
\end{figure}

\begin{figure}[b]
\begin{center}
\epsfig{bbllx=2.0truecm,bblly=9.4truecm,bburx=20truecm,bbury=19.4truecm,clip=,%
width=18truecm,file=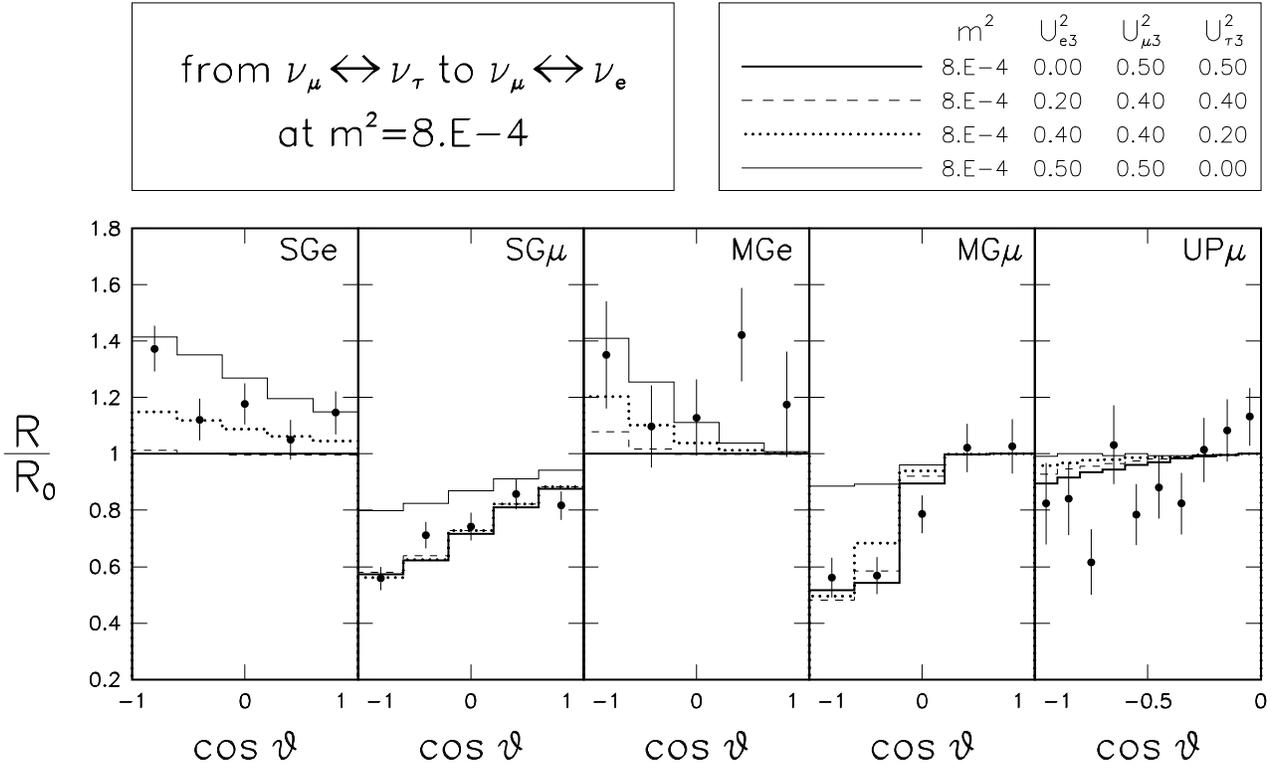}
\caption{As in Fig.~11, but for $m^2=8\times 10^{-4}$ eV$^2$ (i.e., below
CHOOZ bounds).}
\end{center}
\end{figure}

\subsection{Fit to Super-Kamiokande data}

Figure~14 shows our three-flavor fit to the SK data in the triangle  graph,
for values of $m^2$ decreasing from $2.5\times 10^{-2}$ to $4.0\times
10^{-3}$ eV$^2$. The curves at 90\% and 99\% C.L.\ correspond to  an increase
of $\chi^2$ by 6.25 and 11.36 above the global minimum. The  results, 
shown separately for sub-GeV, multi-GeV, and upward-going muons  in the first
three columns of triangles, are then combined in the  last column.
The CHOOZ data are excluded in this fit, in order to study  what one can learn
just from the SK data.

For $m^2=2.5\times 10^{-2}$ eV$^2$, the sub-GeV data exclude all two-flavor
oscillation subcases (the triangle sides), but are consistent with genuine
three-flavor oscillations at large $U^2_{e3}$ ($\gtrsim 0.5$). Also multi-GeV
data are not in agreement with two-flavor oscillations, although the
$\nu_\mu\leftrightarrow\nu_\tau$ subcase (lower side of the  triangle) cannot
be excluded at 99\% C.L. Multi-GeV data are well fitted by genuine
three-flavor oscillations, but in a range of $U^2_{e3}$ different  (lower)
than for sub-GeV data.  The quality of the MG fits improves rapidly  as one
moves from the right side ($\nu_\mu\leftrightarrow\nu_e$)  to the inner part
of the triangle, as expected from the discussion of Fig.~11 in Sec.~V~A. The
good $3\nu$ fit to SG and MG data is mainly driven by the genuinely $3\nu$
matter effects discussed in Sec.~V~A [see Eq.~(\ref{dP}) and related
comments]. Upward going muon data are much less constraining---at 99\% C.L.\
they allow any oscillation scenario. At 90\% C.L.\ they disfavor:  (i) Pure
or quasi-pure $\nu_e\leftrightarrow\nu_\tau$ and $\nu_\mu\to\nu_e$ 
oscillations (left and right sides); and (ii) Large three-flavor mixing (the
central region of the triangle). Large $3\nu$ mixing is excluded because  it
suppresses and distorts too much  the UP$\mu$ distribution (see Fig.~11).
The regions allowed separately by SG, MG, and UP data have no common
intersection, and the combination of all the data is a null region (last
triangle).

For $m^2=1.5\times 10^{-2}$ eV$^2$, the increasing  energy-angle dependence
of the oscillation probability helps to fit the data better.  
Therefore, the regions
allowed in the triangle are larger for each of the  three data sets, and a
small allowed region appears at 99\% C.L.\ in the  combination. Such region
is enlarged for a lower value of $m^2$ ($1.0\times  10^{-2}$ eV$^2$) and it
appears also at 90\% C.L.\ at  $m^2=6.5\times 10^{-3}$ eV$^2$. For the latter
value of $m^2$, all the data  are consistent with
$\nu_\mu\leftrightarrow\nu_\tau$ oscillations (the lower  side), while the
other two-flavor subcases (the left and right sides) are  excluded. The MG
data sample  ``repels'' $\nu_\mu\leftrightarrow\nu_e$  oscillations more
strongly than SG data since, as observed in Sec.~II~A,  higher-energy samples
are characterized by a larger $\mu/e$ unoscillated  ratio, and thus are more
sensitive to the presence (or absence) of  $\nu_\mu$-$\nu_e$ mixing.

For $m^2\sim 4.0\times 10^{-3}$ eV$^2$, also the global 
combination of the data is
consistent with $\nu_\mu\leftrightarrow\nu_\tau$ oscillations, with large but
not necessarily maximal mixing. Large values of $U^2_{e3}$ are also allowed,
indicating that the SK data, by themselves, do not exclude large three-flavor
mixing. We remark that the goodness of the fit improves rapidly when one moves
from the right side inwards  (i.e, when one ``perturbs'' pure
$\nu_\mu\leftrightarrow\nu_e$ oscillations), while it
changes more slowly when one moves
from the lower side upwards (i.e., when one  ``perturbs'' pure
$\nu_\mu\leftrightarrow\nu_\tau$ oscillations), as  expected from the
discussion in Sec.~V~A.

Figure~15 is analogous to Fig.~14, but for lower values of $m^2$, ranging
from $2.5\times 10^{-3}$ to $4.0\times 10^{-4}$ eV$^2$. As $m^2$ decreases,
the SG data fit is not affected very much, since values as low as few$\times
10^{-4}$ eV$^2$ still provide a good fit  to this sample. However, the
oscillation  phase starts decreasing more  rapidly for higher-energy samples
(MG and UP), leading to an insufficient  suppression of the muon rates and to
a gradual reduction of the allowed  regions.  Notice, in particular,
how UP$\mu$ data constrain $3\nu$ mixing for $m^2\sim 1$--3 eV$^2$. 
In any case, the preferred
regions are more and more reduced 
and closer to the lower side of the triangle, corresponding to smaller
allowed values for $U^2_{e3}$. There is no joint allowed region at 99\% C.L.\ 
for $m^2$ below $\sim 5\times 10^{-4}$ eV$^2$.

We summarize the three-flavor fit to  SK data as follows: (i) The  SK data
exclude both $U^2_{\mu3}=0$ and $U^2_{\tau 3}=0$, being consistent  with
large $\nu_\mu\leftrightarrow\nu_\tau$ mixing 
(not necessarily maximal); (ii)  Values of
$U^2_{e3}$ as large as 0.5 cannot be excluded only on the basis  of SK data.
Indeed, $3\nu$ oscillations with large $\nu_e$ mixing can improve the fit
and, in particular, it can explain (part of) the electron excess in the SG
and MG samples. Quantitative bounds on the mixing matrix elements
$U^2_{\alpha 3}$ can be derived from Figs.~14 and 15. In the next subsection we
study the impact of CHOOZ on  such indications.

\begin{figure}[b]
\begin{center}
\epsfig{bbllx=1.0truecm,bblly=4.5truecm,bburx=20truecm,bbury=26.5truecm,clip=,%
width=18truecm,file=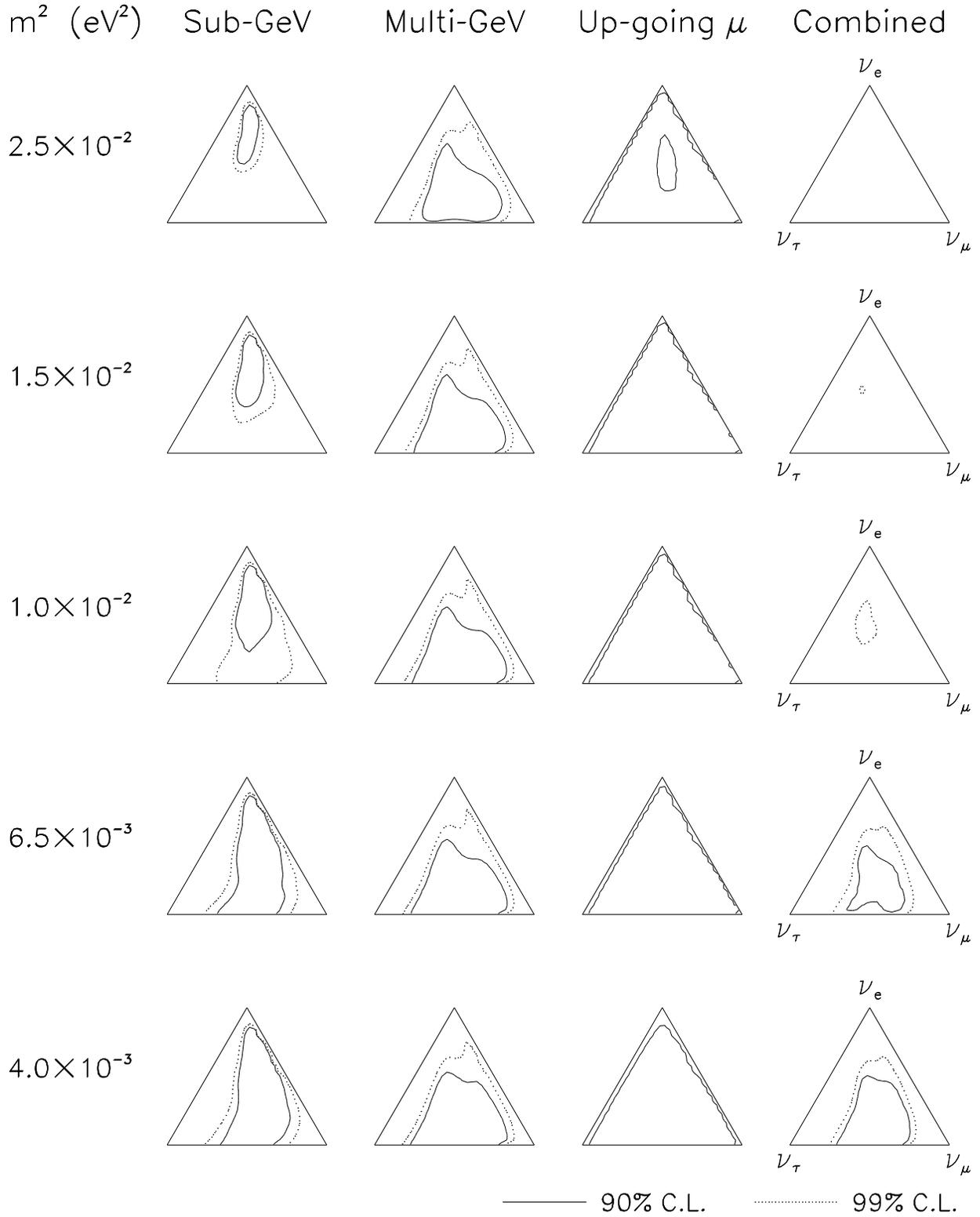}
\caption{Three-flavor analysis in the triangle plot.  Separate and combined
fits to the Super-Kamiokande atmospheric  neutrino data, for selected values
of $m^2$. The parameter space is defined in Fig.~4. Notice that the allowed
regions for the combined fit are always close to the lower side of the
triangle, i.e., to pure $\nu_\mu\leftrightarrow\nu_\tau$ oscillations,
although it is not necessarily so for the separate data samples. Relatively
large values of $\nu_e$ mixing $(U^2_{e3})$ are allowed. }
\end{center}
\end{figure}

\begin{figure}[b]
\begin{center}
\epsfig{bbllx=1.0truecm,bblly=4.5truecm,bburx=20truecm,bbury=26.5truecm,clip=,%
width=18truecm,file=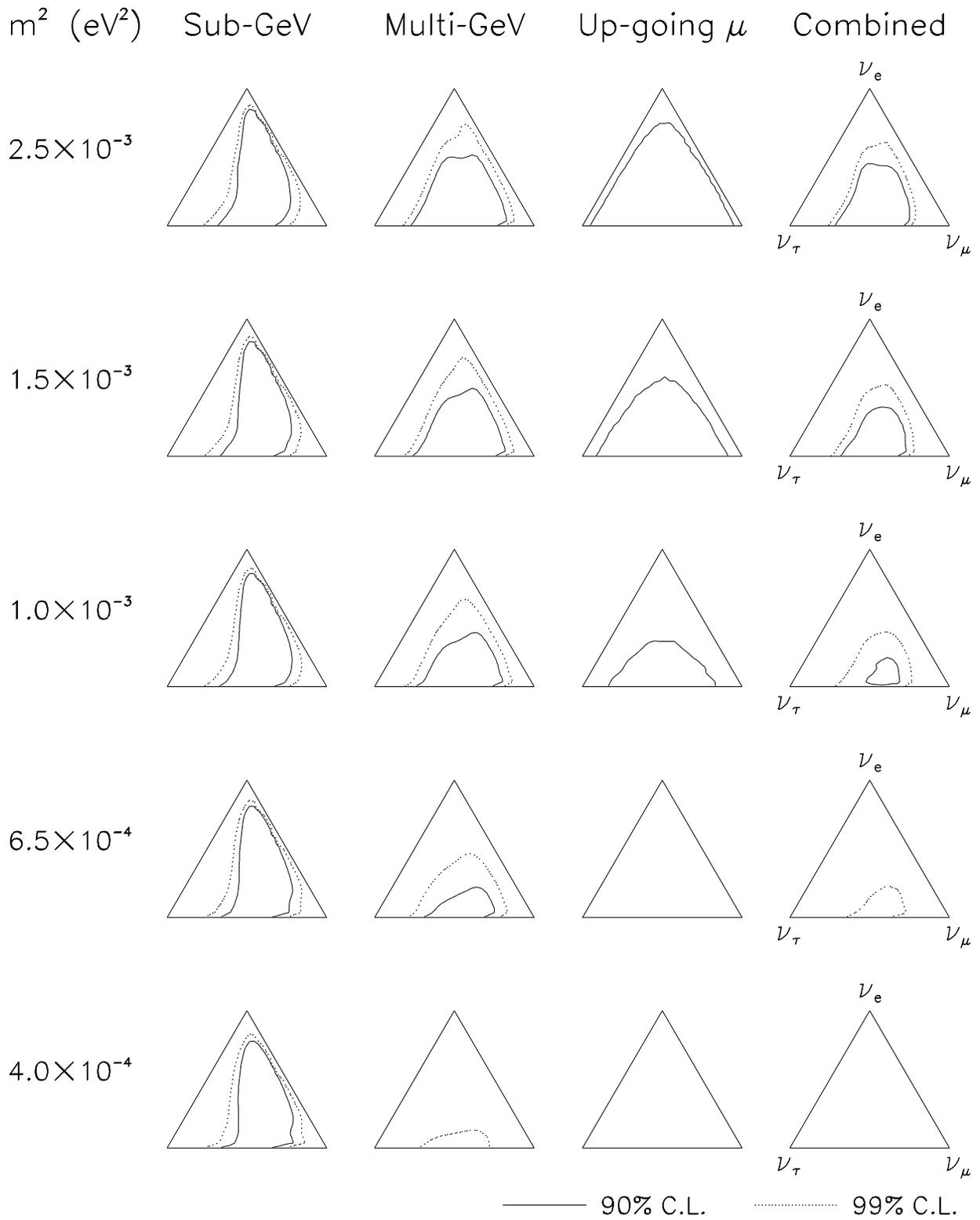}
\caption{As in Fig.~14, but for different (lower) values of $m^2$. }
\end{center}
\end{figure}

\subsection{Fit to Super-Kamiokande and CHOOZ data}

We combine SK and CHOOZ data (30+1 observables) through a joint $\chi^2$
analysis. The results are reported in Fig.~16.

Figure~16 shows the 90\% and 99\% C.L.\ in the triangle plot for selected
values of $m^2$, ranging from $4\times 10^{-3}$ eV$^2$ (upper triangles) to
$6.5\times 10^{-4}$ eV$^2$ (lower triangles). The left column of triangles
reports the fit to SK data only (as  derived from Figs.~14 and 15). The
middle column reports the fit to CHOOZ  data, which exclude a large
horizontal stripe. In fact, the nonobservation of $\nu_e$ disappearance
implies that $\nu_3$ is either very close to the upper corner $\nu_e$ (so as
to suppress oscillations) or very close to the lower side
($\nu_\mu\leftrightarrow\nu_\tau$ oscillations being unobservables in CHOOZ).
Clearly, the addition of the CHOOZ bounds to the SK fit (right column of
triangles) cuts significantly the upper part of the solutions, so that only
relatively low values of $U^2_{e3}$ are allowed. However, the CHOOZ bounds
are rapidly weakened as $m^2$ is decreased, and for $m^2=1.5\times 10^{-3}$
eV$^2$  the parameter $U^2_{e3}$ can be as large as $\sim 0.15$ at 90\% C.L.\
and  $\sim 0.25$ at 99\% C.L., corresponding to a significant $\nu_e$
appearance  probability [see Eq.~(\ref{P})]. Therefore, the CHOOZ data
constrain but {\em do not\/} exclude the role of $\nu_e$ mixing and electron
appearance in the interpretation of the SK data (see also
\cite{Zeni,Foot}). In particular, part of the electron  excess in the SG and
MG samples could be explained by nonzero values of  $U^2_{e3}$ rather than by
uncertainties in the overall neutrino flux  normalization. Nonzero values of
$U^2_{e3}$ also contribute to distort  the zenith distributions \cite{Zeni},
as discussed in Sec.~V~A.

So far we have seen the impact of CHOOZ on the mixing parameters $U^2_{\alpha
3}$. The impact on the square mass difference $m^2$ is summarized in Fig.~17,
which shows the $\chi^2$ as a function of $m^2$, for unconstrained values of
the mixing angles and for both fits to SK (dashed lines) and SK+CHOOZ (solid
lines). The minimum value of $\chi^2$ is $28.3$ for SK and $29.8$ for
SK+CHOOZ, indicating a good fit to the data (30 and 31 observables,
respectively). The CHOOZ data help to constrain $m^2$ on the higher range,
but its role   decreases rapidly for $m^2\lesssim 10^{-3}$ eV$^2$.

Also shown in Fig.~17 are the 90\% and 99\% C.L.\ intervals for $m^2$, which
allow values as low as  $5\times 10^{-4}$ eV$^2$. The possibility of
exploring such low values of $m^2$ should be seriously considered in long
baseline experiments. An interesting result of Fig.~17 is the stability of
the $m^2$ range indicated by SK---it does not change dramatically by adding the
CHOOZ constraint. Therefore, the inclusion of the CHOOZ data in the
global analysis affects more the mixing than the mass parameter.

A comparison of SK and pre-SK bounds is illuminating. Figure 17 should be
compared with Fig.~10 of \cite{3ATM}, where we combined the data from NUSEX,
Fr{\'e}jus, IMB, and Kamiokande (sub-GeV and multi-GeV) in a three-flavor
analysis. The comparison shows that the SK and SK+CHOOZ bounds on $m^2$ are
perfectly consistent with the pre-SK bounds.  The SK+CHOOZ data appear to
improve significantly the old upper bound on $m^2$, but give a lower bound very
similar to the pre-SK data. Notice that we have long since claimed that the
popular value  $m^2\sim 10^{-2}$ eV$^2$ overestimated the best fit for pre-SK
data, and that values as low as $5\times 10^{-4}$ eV$^2$ were compatible with
the atmospheric $\nu$ data \cite{3ATM}.  We plan to perform a joint analysis
of all the data  (SK+CHOOZ+pre-SK) in a future work.

\section{Implications of the $3\nu$ analysis}

In this Section we examine some implications of our three-flavor analysis for
the phenomenology of atmospheric, long-baseline, and solar neutrino
experiments, as well as for model building.

\subsection{Atmospheric $\nu$ phenomenology}

The SK atmospheric data are consistent with $3\nu$ oscillations with dominant
$\nu_\mu\leftrightarrow\nu_\tau$ transitions  ($U^2_{\mu 3}\cdot U^2_{\tau
3}\gtrsim 0.2$ at 90\% C.L.) and subdominant
$\nu_e\leftrightarrow\nu_{\mu,\tau}$ mixing $(U^2_{e3}\lesssim 0.15)$. The
mass square difference $m^2$ is favored in the range $\sim 0.8\times
(10^{-3}$--$10^{-2})$ eV$^2$. Can one improve significantly such indications
only with SK or other atmospheric data?

Large $\nu_\mu\leftrightarrow\nu_\tau$ mixing should generate a $\nu_\tau$
flux comparable to the $\nu_\mu$ flux at the detector site. However, the
``contamination'' of $\mu$-like and $e$-like events from $\tau$
production and subsequent leptonic decay is estimated to be very small in SK
\cite{EVID}. Therefore, there seems to be little hope to test the
$\nu_\mu\leftrightarrow\nu_\tau$ channel through $\tau$ appearance in SK.
Nevertheless, the possibility of enhancing (through appropriate cuts) the
$\tau\to\mu$ and $\tau\to e$ ``pollution'' in selected $\mu$ and $e$ event
samples may deserve further attention in other atmospheric $\nu$ detectors
such as Soudan2.

The tests of $\nu_e$ mixing (i.e., of the matrix element $U^2_{e3}$) can
certainly be improved with higher statistics SK data, in particular with more
multi-GeV upgoing electron and muon events. Such data samples are
characterized by a relatively high $\nu_\mu/\nu_e$ flux  ratio (see
Sec.~II~A), and thus are more sensitive to an increase of $e$-like events due
to $\nu_\mu\leftrightarrow\nu_e$ transitions. Multi-GeV data are already more
powerful than SG (and UP$\mu$) data in constraining $U^2_{e3}$ (see Fig.~15).

\begin{figure}[b]
\begin{center}
\epsfig{bbllx=1.0truecm,bblly=4.5truecm,bburx=20truecm,bbury=26.5truecm,clip=,%
width=18truecm,file=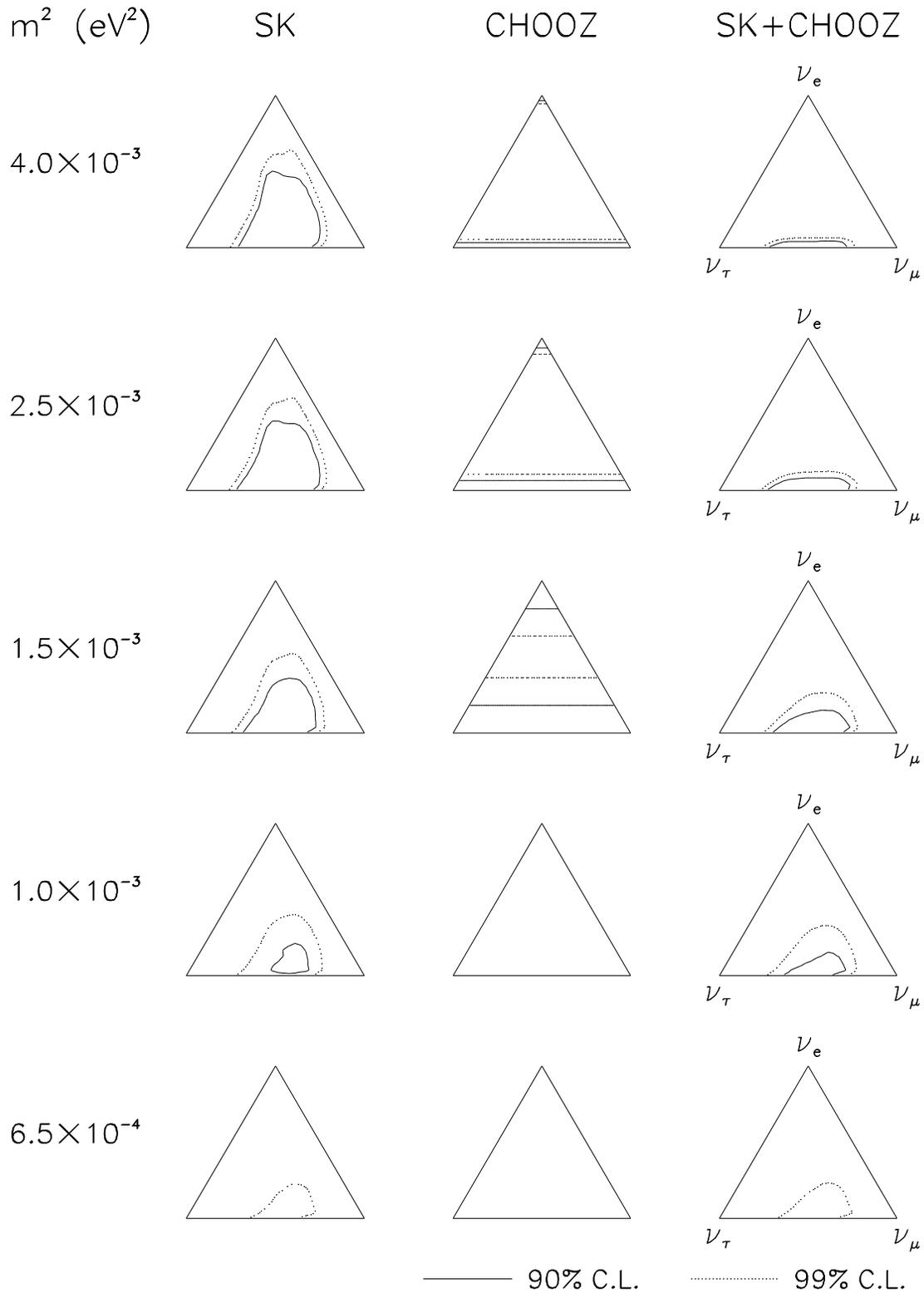}
\caption{Three-flavor analysis in the triangle plot.  Separate and combined
fits to Super-Kamiokande and CHOOZ. The CHOOZ data exclude large horizontal
stripes in the triangle plots. The combined SK+CHOOZ solutions are closer to
pure $\nu_\mu\leftrightarrow\nu_\tau$ oscillations as compared with the fit
to SK data only. However, the allowed values of $U^2_{e3}$ are never
negligible, especially  in the lower range of $m^2$.}
\end{center}
\end{figure}
\begin{figure}[b]
\begin{center}
\epsfig{bbllx=1.0truecm,bblly=8truecm,bburx=20truecm,bbury=24.truecm,clip=,%
width=18truecm,file=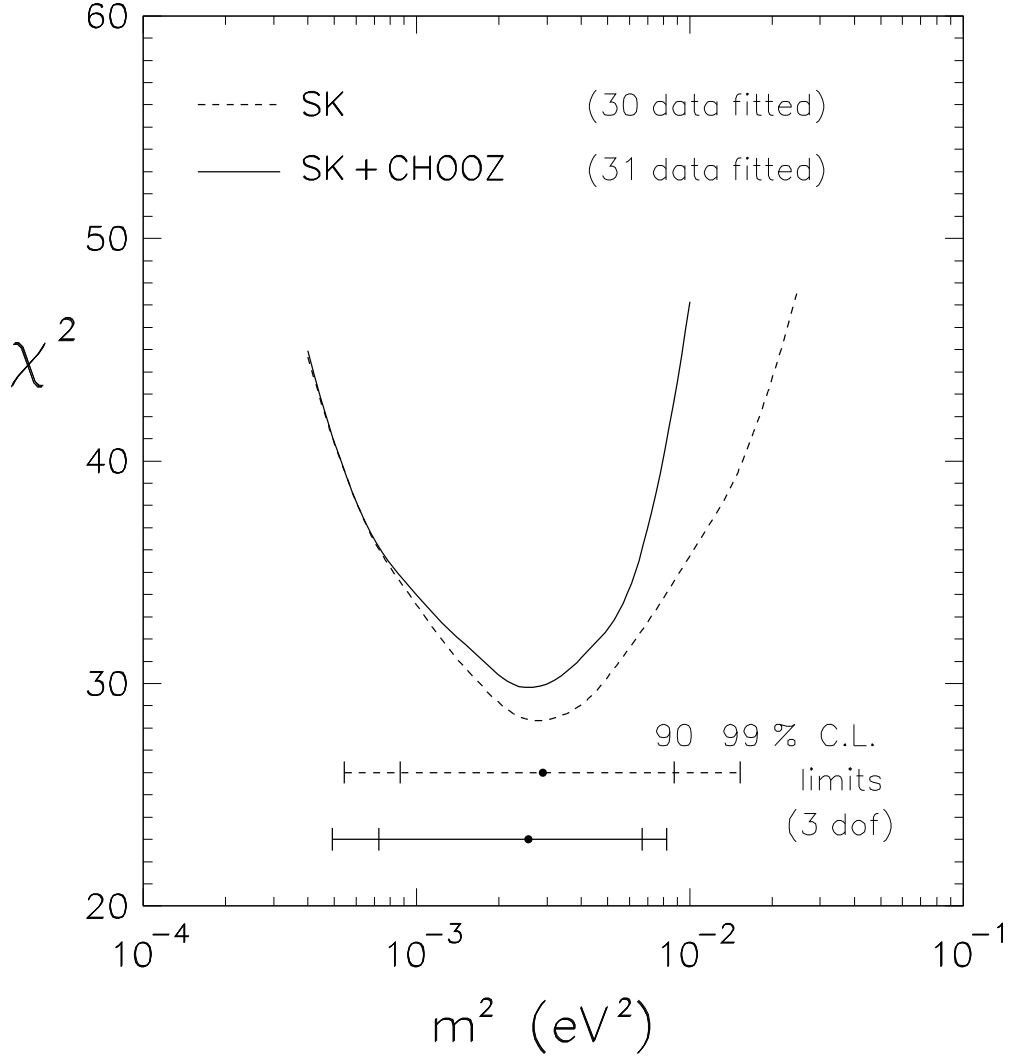}
\caption{Bounds on $m^2$ for unconstrained mixing, as derived from the
$\chi^2$ analysis of SK data, with and without CHOOZ.	 The 90\% and 99\%
C.L.\ intervals correspond to variations of $\chi^2-\chi^2_{\rm min}=
6.25,\,11.36$ for three degrees of freedom (the oscillation parameters).}
\end{center}
\end{figure}

\bigskip\bigskip\bigskip

While the elements $U^2_{\alpha3}$ determine the {\em amplitude\/} of
oscillations, which can be already derived from total event rates, the
parameter $m^2$ governs the {\em phase\/} of oscillations, and thus it can be
derived only through event spectra. Hypothetical spectra of  neutrino events
as a function of $E_\nu$ and $L$ would be the most  sensitive probes of
$m^2$. Unfortunately, a complete kinematical closure of $\nu$-induced events
cannot be achieved in SK, so neither $E_\nu$ nor $L$ can be precisely
reconstructed, especially for low-energy events. This intrinsic feature will
eventually limit the maximum accuracy of $m^2$ fits attainable with SK data
only. In this respect, the  possibility of improving the $E_\nu$ and $L$
reconstruction in experiments as Soudan2 \cite{Soud} (through observation of 
the struck nucleon), or in high-density detectors as proposed in
\cite{Raga}, appears extremely interesting and promising.

\bigskip\bigskip\bigskip

\subsection{Long Baseline experiments}

Long baseline (LBL) accelerator experiments, such as K2K \cite{KtoK}, MINOS
\cite{MINO}, and various CERN-Gran~Sasso proposals \cite{Beam}, are expected
to confirm the atmospheric neutrino signal with a controlled beam. Since both
two-flavor \cite{EVID}  and three-flavor analyses like ours show that $m^2$
can be as low as  $\sim 5\times10^{-4}$ (99\% C.L.), the design of
low-energy  beams should be  pursued seriously. If the atmospheric $m^2$
range can be covered  completely, then it suffices to have either a $\nu_\mu$
disappearance or a $\nu_\tau$ appearance signal to confirm the SK anomaly.

However, we think that long baseline experiments should be designed to {\em
measure\/} oscillation parameters, rather than merely {\em to confirm}  an
oscillation effect already found by SK. 
Measuring the oscillation parameters is a task that demands careful
considerations, especially  if $3\nu$ oscillations are to be tested (for
$2\nu$ oscillations, see the lucid discussion in \cite{BaLi}).

The determination of $m^2$ requires energy spectra analyses, and thus
high-statistics event samples.  If $m^2$ happens to be in the low range
of the experimental sensitivity, the $\tau$ appearance sample might
consist of just a handful of events. The $e$ appearance event sample might
also be small if $U^2_{e3}\to 0$. Therefore, a safe reconstruction of $m^2$ 
should be based mainly on $\mu$ event spectra
from the $\nu_\mu\to\nu_\mu$ disappearance channel, where most of the signal
is expected in any case. This implies a good monitoring of the initial
$\nu_\mu$ beam with a near detector.

The determination of the matrix elements $U^2_{\alpha3}$ requires that several
oscillation channels are probed at the same time---redundancy is never enough
to constrain neutrino mixing \cite{LBLE}. For instance, the
$\nu_\mu\to\nu_\mu$ disappearance channel is sensitive only to $U^2_{\mu 3}$
but tells nothing on $U^2_{e 3}$ or $U^2_{\tau 3}$, while the $\nu_\mu
\to\nu_\tau$ appearance channel is sensitive to the product $U^2_{\mu
3}U^2_{\tau 3}$ but it cannot separate the two factors $ U^2_{\mu 3}$ and $
U^2_{\tau 3}$ nor measure $U^2_{e3}$ [see Eqs.~(\ref{P1},\ref{P})]. These
aspects of $3\nu$ mixing tests \cite{AcRe,LBLE} in long-baseline experiments 
are better appreciated in Fig.~18.

Figure 18 shows, in the first column of  triangles, the region allowed by
SK+CHOOZ for selected values of $m^2$. The second column shows, superimposed,
the prospective regions that can be probed at $\sim90\%$ C.L.\ by K2K 
\cite{KtoK}, both
in the $\nu_\mu\to\nu_\mu$ channel (slanted bands)  and in the
$\nu_\mu\to\nu_e$ channel (hyperbola).%
\footnote{Curves of isoprobability are either of the form
	$U^2_{\alpha 3}={\rm const}$ in the $\nu_\alpha\to\nu_\alpha$ 
	disappearance channel or of the form $U^2_{\alpha 3}U^2_{\beta 3}=
	{\rm const}$ in the $\nu_\alpha\to\nu_\beta$ appearance channel 
	\protect\cite{AcRe}.}
It appears that K2K might not reach, in the $\nu_\mu\to\nu_e$ channel,
sufficient sensitivity to probe the values of $U^2_{e3}$ allowed by SK+CHOOZ.
The disappearance channel $\nu_\mu\to\nu_\mu$ can cover the whole SK+CHOOZ
region, but only for $m^2\gtrsim 2\times 10^{-3}$ eV$^2$. Therefore, K2K is
basically expected to give information on $U^2_{\mu 3}$ for  $m^2\gtrsim
2\times 10^{-3}$ eV$^2$, given the sensitivities prospected in \cite{KtoK}.

With respect to K2K, the MINOS experiment is being designed to probe lower
values of $m^2$ and to explore 
also the $\nu_\mu\to\nu_\tau$ appearance channel (the
region below  the hyperbola touching the $\nu_\mu\to\nu_\tau$ side in
Fig.~18). Possible signals in the three channels $\nu_\mu\to\nu_{e,\mu,\tau}$
will constrain the quantities $U^2_{\mu3}U^2_{e3}$, $U^2_{\mu 3}$, and
$U^2_{\mu3}U^2_{\tau3}$, respectively, so that the elements $U^2_{\alpha 3}$
can be pinpointed for $m^2 \gtrsim 2\times 10^{-3}$ eV$^2$
if the uncertainties are
kept small. For lower $m^2$'s, MINOS rapidly looses sensitivity in at least
one of the oscillation channels,  and it might be difficult to constrain the
neutrino mixing parameters.

Notice that, for $U^2_{e3}\sim 0.15$ and the $U^2_{\mu 3}\sim 0.5$ (allowed
by SK+CHOOZ), the $\nu_\mu\to\nu_e$ appearance probability is $P_{\mu
e}=4U^2_{\mu3}U^2_{e3}\langle S\rangle \sim 0.3\cdot \langle S \rangle$,
where $\langle S \rangle(<1)$ is the oscillation factor  in Eq.~(\ref{S}),
averaged over the $\nu$ beam energy spectrum. Depending on 
$\langle S\rangle$ and on the specific mixing parameters, values of
$P_{\mu e}$ as large as 15\% appear possible in properly designed LBL
experiments.

A final remark is in order. The sensitivity regions in Fig.~18 have been
derived from the prospective estimates reported in the experiment proposals
\cite{KtoK,MINO}, which are in continuous evolution (even more so for the
CERN to Gran Sasso proposals \cite{Beam}, not shown). Therefore, the above
considerations on K2K and MINOS are to be considered as preliminary and
qualitative. Nevertheless, it remains true that LBL experiments might face
some difficulties in constraining the $3\nu$ mixing parameters, especially if
$m^2$ is low or if the three oscillation channels
$\nu_\mu\to\nu_{e,\mu,\tau}$ cannot all be probed.  Re-directing  the goal of
LBL experiments from ``confirming the Super-Kamiokande signal'' to
``measuring  the parameters $(m^2,U^2_{e3},U^2_{\mu3},U^2_{\tau 3})$'' would 
be beneficial to the current debate on various LBL proposals.

\begin{figure}[b]
\begin{center}
\epsfig{bbllx=1.0truecm,bblly=4.5truecm,bburx=20truecm,bbury=26.5truecm,clip=,%
width=18truecm,file=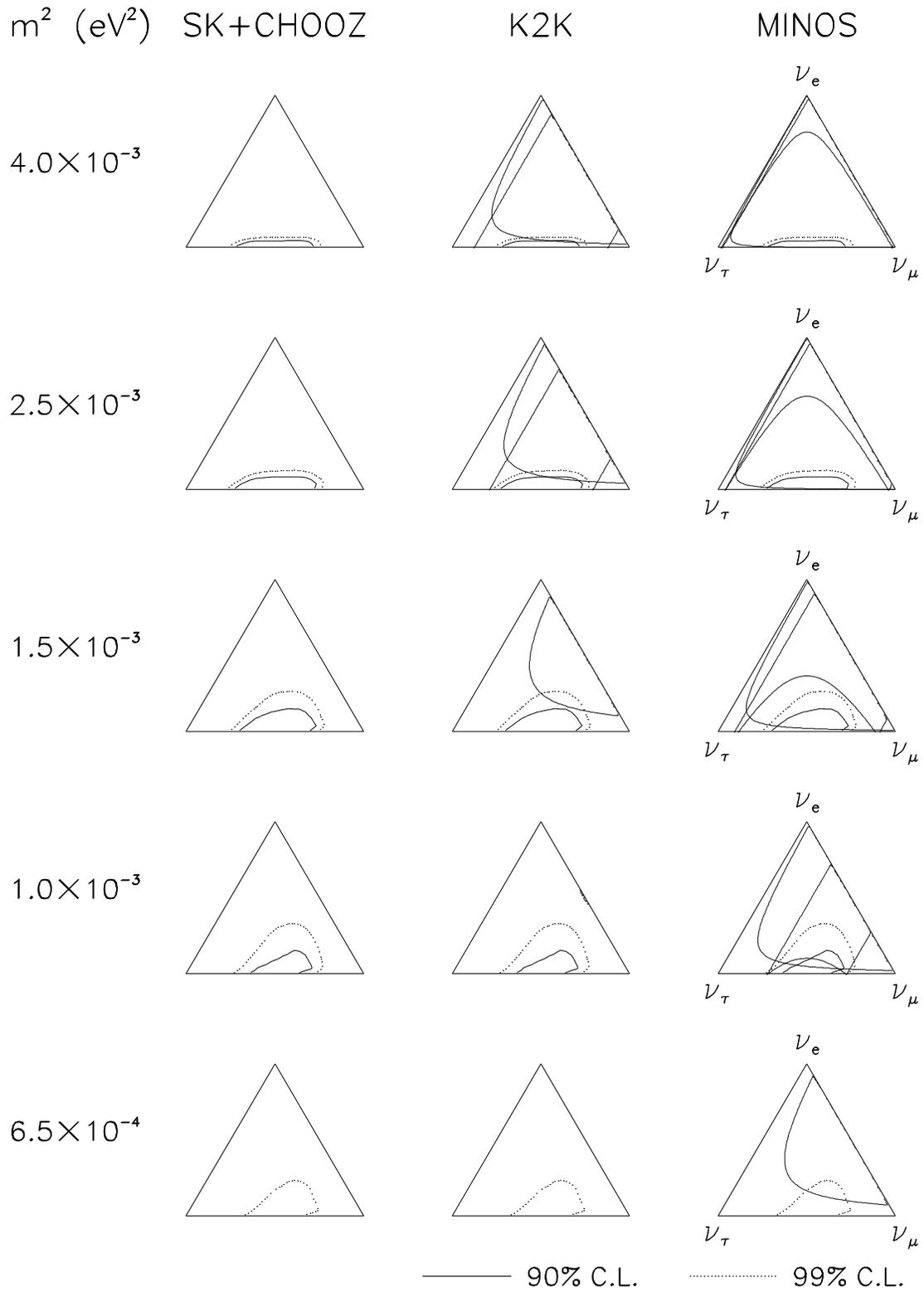}
\caption{Regions of the SK+CHOOZ solutions explorable by two Long Baseline
accelerator experiments (K2K and MINOS) through various oscillation channels.
See the text for details.}
\end{center}
\end{figure}

\begin{figure}[b]
\begin{center}
\epsfig{bbllx=2.0truecm,bblly=9.4truecm,bburx=20truecm,bbury=19.4truecm,clip=,%
width=18truecm,file=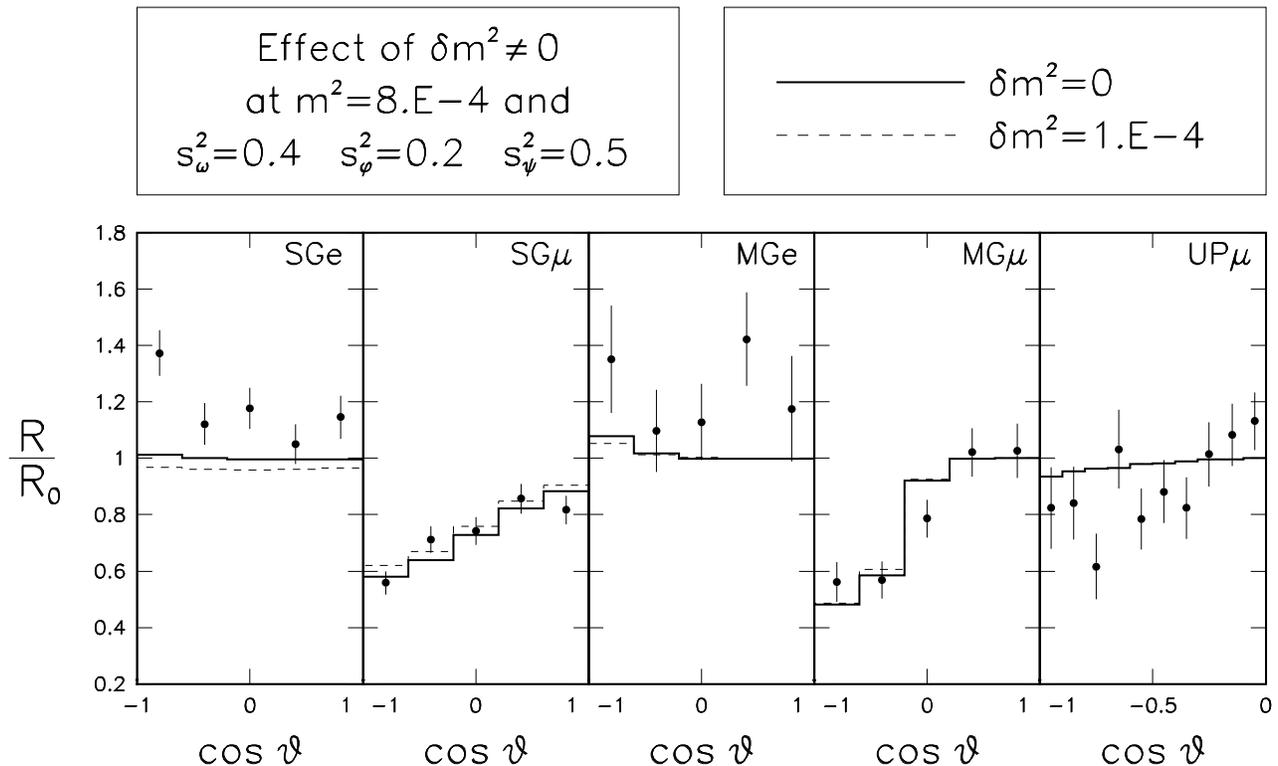}
\caption{Beyond the approximation of one mass scale dominance: example of
effects induced by the subdominant mass scale $\delta m^2$, for oscillation
parameters allowed by atmospheric, CHOOZ, and solar neutrino data.}
\end{center}
\end{figure}

\subsection{Solar neutrino problem}

In the limit $\delta m^2\ll m^2$  [Eq.~(\ref{appr})], experiments with
terrestrial  (atmospheric, accelerator and reactor) neutrino beams probe
the parameters $(m^2,U^2_{e3},U^2_{\mu3},U^2_{\tau3})$. In the same
approximation, solar neutrino experiments probe the parameters $(\delta
m^2,U^2_{e1},U^2_{e2},U^2_{e3})$, i.e.,  the small mass square difference and
the mass composition of $\nu_e$ \cite{3MSW}.

Therefore, both terrestrial and solar $\nu$ experiments probe the element
$U^2_{e3}$, which parametrizes the mixing of $\nu_e$ with the ``lone''
neutrino mass eigenstate $\nu_3$. The effect of such mixing on solar
neutrinos is to give an energy-independent contribution to the disappearance
of  $\nu_e$'s from the sun. Since this contribution is basically proportional
to the square of $U^2_{e3}$ \cite{3MSW},  sizable mixing is required to have
large effects. However, as $U^2_{e3}$ is constrained by the SK+CHOOZ fit, 
such effects are relatively small to be detected in the current 
solar neutrino experiments.
The smallness of $U^2_{e3}$ also reduces the ``coupling'' of the terrestrial
and solar parameter spaces \cite{CHde}.

Nevertheless, it is interesting to notice that a preference for small values
of $U^2_{e3}$ emerges naturally from solar neutrino data only, in both 
matter-enhanced \cite{Comp,3MSW} and vacuum \cite{3VAC} three-flavor
oscillation fits.  An updated analysis using the latest SK solar neutrino
data would be desirable to confirm such indication.  In fact, a more
pronounced preference of  solar $\nu$'s for relatively small values of
$U^2_{e3}$ would be a nontrivial, although indirect, hint that both solar and
terrestrial $\nu$ data are consistent within the same $3\nu$ oscillation
scenario.

The relation between solar and atmospheric neutrinos is not necessarily
confined to a preference for relatively small values of the mixing matrix
element $U^2_{e3}$. For instance, if the assumption in
Eq.~(\ref{appr}) is violated, atmospheric neutrinos can become sensitive to
the subleading oscillations driven by $\delta m^2$, i.e., by the ``solar
neutrino'' mass difference. Figure~19 shows a representative case beyond the
$\delta m^2\ll m^2$ approximation. The values chosen for the oscillation
parameters $(\delta m^2,m^2,\omega,\phi,\psi)$ are allowed at 99\% C.L.\ by the
present SK+CHOOZ bounds and by the pre-SK, three-flavor solar $\nu$
analysis in \cite{3MSW}. The effect of the subdominant mass scale  $\delta
m^2$ can be evaluated by comparing the solid curves ($\delta m^2=0$) and the
dashed curves ($\delta m^2\neq 0$). Of course, the effect is more significant
when $\delta m^2 L/E_\nu$ is larger, i.e., for upgoing SG events, while it
rapidly decreases at higher energies (MG and UP$\mu$ samples) and shorter
pathlengths ($\cos \theta\to 1$). For SG$\mu$ events, however, the effect
approaches the size of the error bars and thus it might be probed by SK! 
However, the strategies for disentangling the oscillations driven by $m^2$
and $\delta m^2$ are nontrivial and will be investigated in a future work.

\subsection{Models of neutrino mass and mixing}

Theoretical or phenomenological models of neutrino mass and mixing try to
predict, or ``explain,'' the set of parameters $(m_1,m_2,m_3)$ and $U_{\alpha
i}$. Our analysis constrains the subset of parameters $(m^2,U^2_{\alpha 3})$,
provided that $\delta m^2$ is sufficiently small  ($\lesssim 10^{-4}$). Many
models that try to explain solar+atmospheric $\nu$ data fall within this
category, and are thus strongly constrained by the SK+CHOOZ bounds worked out
in this paper. For instance, the so-called bimaximal mixing model \cite{Bima},
characterized by $(U^2_{e3},U^2_{\mu 3}, U^2_{\tau 3})=(0,1/2,1/2)$ for
atmospheric neutrinos [and by $(U^2_{e1},U^2_{e2}, U^2_{e3})=(1/2,1/2,0)$ for
solar neutrinos] is allowed for $m^2\gtrsim 10^{-3}$ eV$^2$ (see Fig.~15).
Conversely, the trimaximal mixing model \cite{Trim}, characterized by 
$(U^2_{e3},U^2_{\mu 3}, U^2_{\tau 3})=(1/3,1/3,1/3)$ for atmospheric neutrinos 
[and by $(U^2_{e1},U^2_{e2},U^2_{e3})=(1/3,1/3,1/3)$ for solar neutrinos]
appears to be strongly disfavored by our combined SK+CHOOZ analysis (it would
correspond to the center of each triangle in Figs.~14 and 15). Of course, many
other models (see, e.g., the classification in \cite{Clas}) can be tested
through our bounds on the oscillation parameters, provided that the dominance
of the mass scale $m^2$ is assumed for atmospheric neutrinos.

Other models try to explain the atmospheric anomaly and the LSND evidence
\cite{Lexp} for $\nu_\mu\to\nu_e$ oscillations, with an allowance for an
energy-averaged suppression of the solar neutrino flux. Arguments disfavoring
such scenarios are discussed in \cite{Sacr} (see, in particular, Table~VI of
\cite{Sacr} and related comments).  One such model has been recently proposed
in  \cite{Thun}, where $m^2\sim 0.4$ eV$^2$ is assumed to drive the
oscillations in the  LSND experiment \cite{Lexp} range, as well as
energy-averaged oscillations of atmospheric $\nu$'s, while $\delta m^2\sim
10^{-4}$--$10^{-3}$ eV$^2$  is assumed to drive energy-dependent oscillations
of atmospheric $\nu$'s. Both $\delta m^2$ and $m^2$ can then contribute to the
solar neutrino deficit through energy-averaged oscillations. Since the bounds
worked out in Sec.~V~D assume $\delta m^2\ll 10^{-4}$ eV$^2$, and thus do not
apply to such model, we have performed a numerical analysis of SK data {\em ad
hoc\/}, using the same mass-mixing parameters  as in  \cite{Thun}.  We find
that the resulting zenith angle distributions of muons (not shown) are only
mildly  distorted, and that the model is disfavored by the SK atmospheric data
at $>99\%$ C.L., with or without matter effects.   We mention that,  for the
choice of parameters in \cite{Thun}, matter effects  influence significantly
the zenith distributions, making them flatter than  in vacuum. The
semi-quantitative calculations in \cite{Thun} showed a more  optimistic
agreement to the SK data, in part because matter effects were  ignored. In
addition, the large value for $U^2_{\tau 3}$ chosen in  \cite{Thun} does not
appear in agreement with the global analysis of laboratory neutrino oscillation
searches (including the LSND data) performed in \cite{LSND}.

\section{Summary and conclusions}

We have performed a three-flavor analysis of the SK atmospheric neutrino
data, in a framework characterized by the mass-mixing parameters
$(m^2,U^2_{e3},U^2_{\mu3},U^2_{\tau 3})$, in the hypothesis
of one mass scale dominance. The variations of the zenith
distributions of $\nu$ events in the presence of flavor oscillations have
been investigated in detail. Fits to the SK data, with and without the
additional CHOOZ data, strongly constrain the parameter space. Detailed
bounds have been shown in triangle graphs, embedding the unitarity
condition $U^2_{e3}+U^2_{\mu3}+U^2_{\tau3}=1$.
The allowed regions include the subcase $U^2_{e3}=0$, corresponding
to pure $\nu_\mu\leftrightarrow\nu_\tau$ oscillations. However, values
of $U^2_{e3}> 0$ are also allowed. In particular,  for
$m^2$ close to (or slightly below) $10^{-3}$ eV$^2$, 
$U^2_{e3}$ can be as large as $\sim 0.15$ (at 90\% C.L.). 
Scenarios with $U^2_{e3}>0$ correspond to genuine three-flavor
oscillations and are characterized by a rich phenomenology,
not only for atmospheric $\nu$'s, but also for solar and laboratory
neutrino oscillation searches. In particular, challenging opportunities
are disclosed for $\nu_e$ appearance searches in long baseline experiments.
Our analysis also 
places strong constraints on models of neutrino mass and mixing.
In addition, we have   
examined many facets of the SK data and of their
interpretation, that will deserve further attention when the experimental
and theoretical uncertainties will be reduced.

\acknowledgments

E.L.\ is grateful to H.\ Minakata and to O.\ Yasuda for fruitful discussions
and for kind hospitality at Tokyo Metropolitan University. 
We thank D.\ Montanino for helpful discussions. The work of A.M.\
and G.S.\ is supported by the Ministero dell'Universit\`a e della Ricerca
Scientifica (Dottorato di Ricerca).

\appendix
\section{Calculation of zenith distributions}

The calculation of the zenith angle distributions of SG and MG
lepton events
involves the numerical evaluation of multiple integrals of the form
\begin{equation}
\Phi\otimes\sigma\otimes\varepsilon\otimes P\ ,
\label{multi}
\end{equation}
where $\Phi$ is the unoscillated neutrino spectrum, $\sigma$ is the
differential cross section for lepton production, 
$\varepsilon$ is the detector efficiency for lepton reconstruction, 
and $P$ is the oscillation probability \cite{Comp,3ATM}.

The efficiency function $\varepsilon$ is not always reported
in the experimental papers. In particular, it has not been explicitly given
by the SK Collaboration so far. We faced a similar problem in the analysis
of the Kamiokande multi-GeV data performed in \cite{3ATM}. Our solution
\cite{3ATM} was to use the energy distribution of the parent neutrinos 
$\nu$ that produce a {\em detected\/} lepton $\ell$, namely,
\begin{equation}
\frac{dN_\nu}{dE_\nu}=\frac{d\Phi_\nu}{dE_\nu}\int\! dE_{\ell}\,
\frac{d\sigma_\nu}{d E_\ell}\, \varepsilon(E_\ell)\ ,
\end{equation}
where $E_\ell$ is the lepton energy. This distribution, which 
gives information on the factor $\Phi\otimes\sigma\otimes\varepsilon$ 
in Eq.~(\ref{multi}), has been published in \cite{Ymul} for the
Kamiokande experiment. Concerning SK, we have used the analogous
information from \cite{Priv}. 

Using the energy distribution of parent neutrinos, it is possible
to reconstruct the zenith distribution of the final leptons, provided
that the smearing induced by neutrino-lepton scattering angle is
taken into account \cite{3ATM}. While for the old Kamiokande multi-GeV
data we approximated this effect with an energy-independent smearing
angle of $\sim 17^\circ$, for SK we properly take into account the
distribution of the lepton scattering angle and its dependence
on the energy, which is especially relevant for SG events
\cite{Conn}. We find good agreement with the SK  estimate of the 
average scattering angle as a function of energy
(as reported in \cite{KaTh}, p.~99).
Concerning the neutrino fluxes, we refer to \cite{Hond} except for
SG events, where we use the 
differential spectra from \cite{Bart} with geomagnetic corrections
\cite{GeBa}.

Since the distributions of parent neutrinos in 
\cite{Priv} are given in arbitrary units, we need to normalize
the total area of our estimated SK lepton distributions (SG and MG,  
in the absence of oscillations) to the corresponding values simulated by 
the SK Collaboration, as reported in Table~I (total rates). For SG events,
this renormalization compensates, in part, for the fact that we use low-energy
$\nu$ fluxes from \cite{Bart,GeBa} instead 
than from \cite{Hond}. Of course, a more direct calculation of
the SG and MG distributions (avoiding the use of indirect information
such as the parent $\nu$ distributions)
is preferable; we intend to perform such calculation when the
SK efficiency function $\varepsilon(E_\ell)$ will be made publicly
available. In any case, our present approach produces results in 
satisfactory agreement with SK zenith distributions for SG and MG
events \cite{Kaji}, as
shown in Fig.~20. The small differences between our calculations
and the SK simulations are not relevant, being comparable
to the SK MonteCarlo statistical error.

Figure~20 also shows the UP$\mu$ distribution, for which we use
a direct computation as in \cite{Marr}, with the following ingredients:
GRV94 DIS structure functions \cite{GRVD}, Lohmann
{\em et al.\/} muon energy losses in the rock \cite{Lohm,HaTh},
and the zenith dependence of the SK
muon  energy threshold from \cite{Post,HaTh}. Also for this distribution,
we obtain a good agreement with the corresponding SK calculation
(with the same inputs).

Notice that Fig.~20 refers to the no oscillation case. Some ``oscillated''
$\mu$-like and $e$-like event
distributions (as well as their ratio $\mu/e$)
have also been presented by the SK Collaboration
in various Conferences, especially for the case of maximal 
$\nu_\mu\leftrightarrow\nu_\tau$ mixing. We obtain good
agreement with SK also in such cases (not shown).

In conclusion, we
are confident that our calculations of the zenith distributions
represent a satisfactory approximation (not a substitute,
of course) of the SK simulations. Improvements 
of our calculations for the SG amd MG samples are possible (with a more
accurate knowledge of the SK detector efficiency)
but do not appear to be decisive at present, in view of the good agreement
reported in  Fig.~20 and of the relatively large theoretical uncertainties
discussed in the following Appendix.

\begin{figure}[b]
\begin{center}
\epsfig{bbllx=2.0truecm,bblly=9.4truecm,bburx=20truecm,bbury=19.4truecm,clip=,%
width=18truecm,file=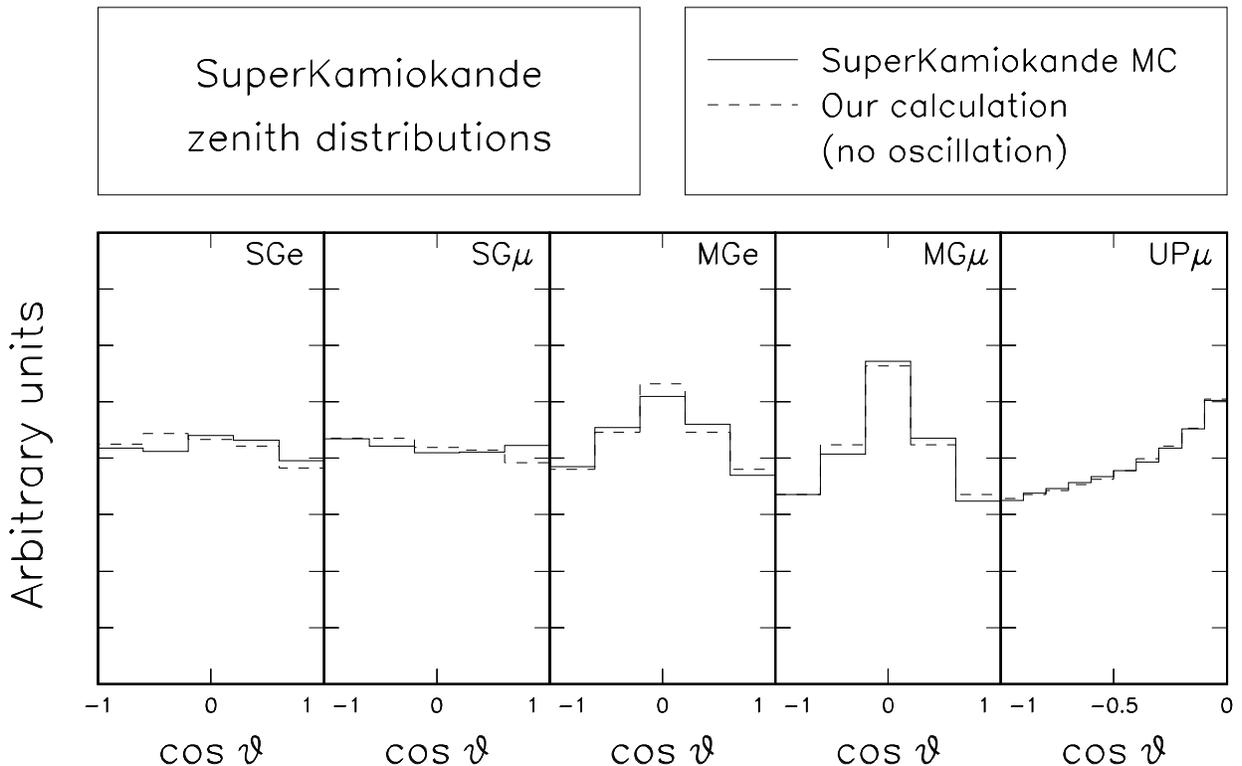}
\caption{Angular distribution shape: comparison of the SK MonteCarlo
simulations (solid lines) with our calculations (dashed lines).}
\end{center}
\end{figure}

\section{Statistical analysis}

Fitting histograms is a delicate task. In general, the predictions
in any two bins are correlated, and ignoring
such correlations typically leads to significant variations
in the allowed ranges for the fit parameters. 
Concerning the SK zenith
distributions, this problem adds to the difficulty of evaluating
the size of uncertainties associated to the theoretical neutrino
flux calculations and to the neutrino cross section, as well as the variations
of such uncertainties in terms of the neutrino energy and direction.

There is no easy solution to such problems, other than continually 
improving the calculations, understanding the role and the uncertainties
of any input parameter, and removing as many approximations
as possible \cite{Gais,Giap,Batt}.  Cosmic ray experiments can also help
to constrain the models of atmospheric showers. The confidence in
the estimated cross sections might benefit from
a resurgence of experimental interest for low-energy neutrino interactions.
In the meantime, it is wise to adopt conservative error estimates.

For each of the 30 SK zenith bins $\{b_i\}_{i=1,\dots,30}$ used in 
the analysis, we take the experimental statistical error $s_i$ 
from Tables~I and II. The $30\times30$ statistical error matrix $s^2_{ij}$
simply reads $s^2_{ij}=\delta_{ij} s_i s_j$. As a global systematic
error $\sigma_i$ for each bin, we assume conservatively $\pm30\%$ of
the theoretical prediction (with or without oscillations). The
systematic error matrix is then $\sigma^2_{ij}=\rho_{ij}\sigma_i\sigma_j$,
where the correlation matrix $\rho_{ij}$ is evaluated as follows.

If the systematic uncertainties $\sigma_i$ had a single, common origin such
as an overall normalization uncertainty, then the bin values $b_i$'s would 
be fully correlated ($\rho_{ij}=1$) and the systematics
would cancel in any bin ratio $b_i/b_j$. However, the presence of
several sources of uncertainties implies that $\rho_{ij}<1$ and that
the ratio $b_i/b_j$ is affected by a residual uncertainty
\begin{equation}
\sigma^2(b_i/b_j) = \sigma^2_i +\sigma^2_j - 2 \rho_{ij} \sigma_i\sigma_j\ ,
\end{equation}
where all $\sigma$'s represent fractional errors. For 
$\sigma_i=\sigma_j=\sigma(=0.3)$ the above relation can be inverted to give
\begin{equation}
\rho_{ij} =1 - \frac{\sigma^2(b_i/b_j)}{2\sigma^2}\ ,
\label{rho}
\end{equation}
which allows to estimate $\rho_{ij}$ from the ratio error.
For instance, if $b_i$ refers to downgoing SG $e$-like events and
$b_j$ to the corresponding $\mu$-like events, with a $\mu/e$ uncertainty
of, say, $\pm 5\%$, the corresponding correlation index is
$\rho_{ij} = 1-(0.05^2)/(2\cdot0.30^2)=0.986$ \cite{Stat}. 

The task is then
reduced to the evaluation of the most important sources of errors
for the ratios $b_i/b_j$. The total error for the $\mu/e$ flavor ratio
(including the theoretical uncertainties and the experimental 
misidentification) is conservatively estimated to be $\pm8\%$ for
SG events and $\pm 12\%$ for MG events in \cite{EVID}. For bins of 
equal flavor, one expects an additional energy-dependent 
uncertainty in the ratio $b_i/b_j$  due to uncertainties in the
neutrino energy spectrum slope. In fact, by comparing
the relative rates of SG, MG, and UP$\mu$ events calculated 
with different input fluxes (either \cite{Bart} or \cite{Hond}), we find
typical ratio errors of $\pm 5\%$ for 
$b_i({\rm SG})/b_j({\rm MG})$ and $b_i({\rm MG})/b_j({\rm UP})$, 
and of $\pm 10\%$ for $b_i({\rm SG})/b_j({\rm UP})$, i.e., errors increasing
with the relative difference between the mean energies of the event samples.
Finally, one expects also angular-dependent errors for $b_i/b_j$, that
we estimate to be at most $\pm 5\%$ when the difference between
$|\cos\theta_i|$ and $|\cos\theta_j|$ is maximal (ratio of vertical to
horizontal direction bins).

Qualitatively, all this means that the correlation between any two
bins decreases from unity
as the bins are more separated in energy, angle, and flavor,
thus giving to the theoretical distributions some freedom to vary 
their shape. Quantitatively, we formalize the above estimates by
generalizing Eq.~(\ref{rho}) as
\begin{equation}
\rho_{ij}=1-
\frac{\sigma^2_f}{2\sigma^2}-
\frac{\sigma^2_E}{2\sigma^2}-
\frac{\sigma^2_\theta}{2\sigma^2}\ ,
\end{equation}
where $\sigma=30\%$, and: 
(i) $\sigma_f$ is the ``flavor-dependent uncertainty,'' equal to
10\% for bins of different flavors and zero otherwise; (ii)
$\sigma_E$ is the ``energy-dependent uncertainty,'' equal to
zero for bins $(i,j)$ belonging to the same sample (SG, MG, or UP), to
5\% for bins $(i,j)$ of the kind (SG,MG) or (MG,UP), and
to 10\% for bins $(i,j)$ of the kind (SG,UP); and (iii)
$\sigma_\theta$ is the ``direction-dependent uncertainty,''
equal to $5\%$ times the difference between the mean direction cosines
$|\langle\cos\theta\rangle_i|$
and $|\langle\cos\theta\rangle_j|$. 
For instance, the first bin of the SG$e$ distribution
and the last bin of the UP$\mu$ distribution have the lowest correlation,
$\rho_{ij}\simeq0.874$, since they are the most distant in energy, flavor,
and direction.

We finally define our $\chi^2$ function as
\begin{equation}
\chi^2 = \sum_{ij}\Delta b_i
\,(s^2_{ij}+\rho_{ij}\sigma_i\sigma_j)^{-1}
\, \Delta b_j\ ,
\end{equation}
where $\Delta b$ is the difference between the bin contents in Tables~I and II
and our theoretical calculations (with or without oscillations).
We mention that the $\chi^2$ fit to the SK data appears to be rather sensitive 
to $\sigma_f$. 
Lowering its value from our present choice (10\%, comparable to the 
estimates in  Table~II of \cite{EVID}) to a few percent would shrink 
significantly  the allowed regions but would also worsen the best fit. 
A reduction of this and other systematics would greatly improve the
statistical power of oscillation hypothesis tests.




\end{document}